\documentclass[reprint, amsmath,amssymb, aps, graphicx]{revtex4-2}
\usepackage[utf8]{inputenc}
\usepackage{times}
\usepackage{amsmath}
\usepackage{braket}
\usepackage{hyperref}
\usepackage{rotating}
\usepackage{amsmath}
\usepackage{geometry}
\usepackage{xfrac}
\usepackage{graphicx}
\usepackage{physics}
\usepackage[modulo,left]{lineno}

\newcommand{\Cteen}{${}^{13}$C}
\newcommand{\FigLet}[1]{(\textbf{{#1}})}

\begin{document}

\title[Qubit teleportation between non-neighboring nodes in a quantum network]{Qubit teleportation between non-neighboring nodes in a quantum network}

\author{S.L.N. Hermans}
\thanks{These authors contributed equally to this work.}%
\author{M. Pompili}
\thanks{These authors contributed equally to this work.}%
\author{H.K.C. Beukers}
\author{S.  Baier}
\altaffiliation[Present address: ]{Institut für Experimentalphysik, Universit{\"a}t Innsbruck, Technikerstraße 25, 6020 Innsbruck, Austria}
\author{J. Borregaard}
\author{R. Hanson}\email{R.Hanson@tudelft.nl}

\affiliation{QuTech and Kavli Institute of Nanoscience, Delft University of Technology, 2628 CJ, Delft, The Netherlands}

\begin{abstract}
	Future quantum internet applications will derive their power from the ability to share quantum information across the network.  Quantum teleportation allows for the reliable transfer of quantum information between distant nodes, even in the presence of highly lossy network connections. While many experimental demonstrations have been performed on different quantum network platforms, moving beyond directly connected nodes has so far been hindered by the demanding requirements on the pre-shared remote entanglement, joint qubit readout and coherence times.
	Here we realize quantum teleportation between remote, non-neighboring nodes in a quantum network. The network employs three optically connected nodes based on solid-state spin qubits. The teleporter is prepared by establishing remote entanglement on the two links, followed by entanglement swapping on the middle node and storage in a memory qubit. We demonstrate that once successful preparation of the teleporter is heralded, arbitrary qubit states can be teleported with fidelity above the classical bound, even with unit efficiency. These results are enabled by key innovations in the qubit readout procedure, active memory qubit protection during entanglement generation and tailored heralding that reduces remote entanglement infidelities.
	Our work demonstrates a prime building block for future quantum networks and opens the door to exploring teleportation-based multi-node protocols and applications.
\end{abstract}
\maketitle

\subsection*{Introduction}
Quantum teleportation is the central routine for reliably sending qubits across lossy network links~\cite{bennett_teleporting_1993} as well as a key primitive of quantum network protocols and applications~\cite{wehner_quantum_2018, Ben-Or2006, Arora2019}. Using a teleporter in the form of a pre-shared entangled state between the sending node and the receiving node, the quantum information is transferred by performing a joint Bell-state measurement on the sender’s part of the entangled state and the qubit state to be teleported, followed by a gate operation on the receiving node conditioned on the Bell-state measurement outcome~\cite{bennett_teleporting_1993}. Since the quantum information is not transmitted by a physical carrier, the protocol is insensitive to loss in the connecting photonic channels and on intermediate nodes. A deterministic Bell-state measurement combined with real-time feed-forward enables unconditional teleportation, in which state transfer is achieved each time a qubit state is inserted into the teleporter.

Pioneering explorations of quantum teleportation protocols were performed using photonic states~\cite{bouwmeester_experimental_1997,boschi_experimental_1998,furusawa_unconditional_1998}. Following the development of quantum network nodes with stationary qubits, remote qubit teleportation was realized between trapped ions~\cite{olmschenk_quantum_2009}, trapped atoms~\cite{nolleke_efficient_2013,langenfeld_quantum_2021}, diamond NV centers~\cite{pfaff_unconditional_2014} and memory nodes based on atomic ensembles~\cite{bao_quantum_2012}.

While future quantum network applications will widely employ teleportation between non-connected nodes in the network, the demanding set of requirements on the pre-shared entanglement, the Bell-state measurement and the coherence times for enabling real-time feed-forward  has so far prevented the realization of teleportation beyond directly connected stationary network nodes.

\begin{figure*}[ht]
    \centering
    \includegraphics{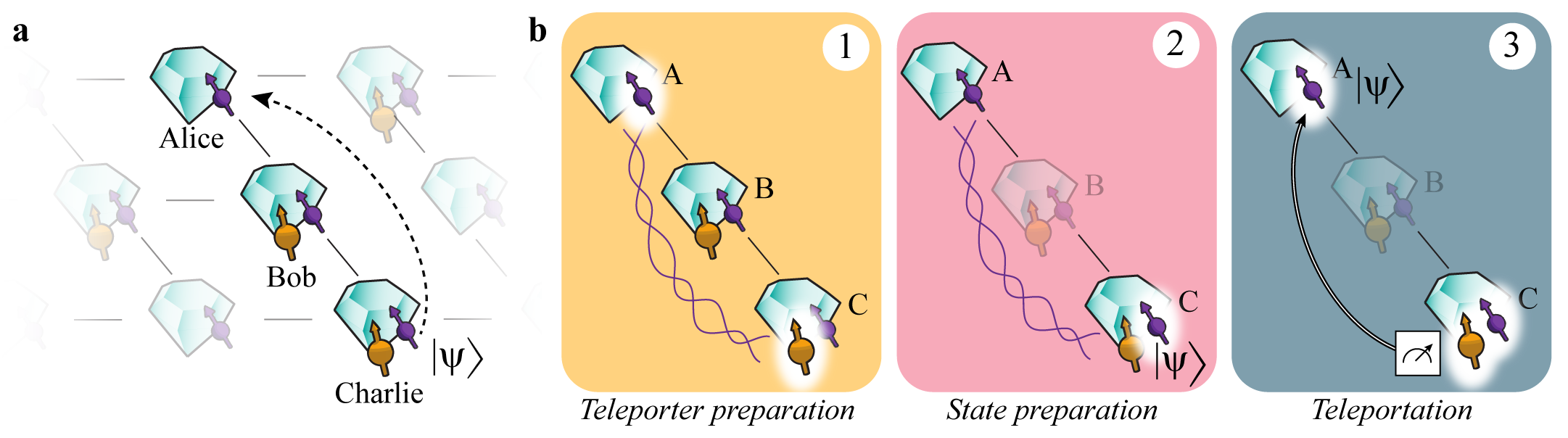}
    \caption{\textbf{Teleporting a qubit between non-neighboring nodes of a quantum network.} 
    \FigLet{a} Three network nodes, Alice (A), Bob (B) and Charlie (C) are connected via optical fiber links (lines) in a line configuration. Each setup has a communication qubit (purple) that enables entanglement generation with its neighboring node. Additionally, Bob and Charlie contain a memory qubit (yellow). 
    \FigLet{b} The steps of the teleportation protocol: (1) We prepare the teleporter by establishing entanglement between Alice and Charlie using an entanglement swapping protocol on Bob, followed by swapping the state at Charlie to the memory qubit. (2) The qubit state to be teleported is prepared on the communication qubit on Charlie. (3) A Bell-state measurement is performed on Charlie’s qubits and the outcome is communicated to Alice over a classical channel. Dependent on this outcome, Alice applies a quantum gate  to obtain the teleported qubit state.}
    \label{fig:teleport_one}
\end{figure*}

Here, we overcome these challenges by a set of key innovations and achieve qubit teleportation between non-neighboring network nodes (see Figure \ref{fig:teleport_one}a). Our quantum network consists of three nodes in a line configuration, Alice, Bob and Charlie. Each node contains a nitrogen-vacancy (NV) center in diamond. Using the NV electronic spin as the communication qubit we are able to generate remote entanglement between each pair of neighboring nodes. In addition, Bob and Charlie each employ a nearby \Cteen$ $ nuclear spin as a memory qubit. The steps of the teleportation protocol are shown in Figure \ref{fig:teleport_one}b. To prepare the teleporter we use an entanglement swapping protocol mediated by Bob to establish entanglement between Alice and Charlie. Once successful preparation of teleporter is heralded, the input qubit state is prepared on Charlie and finally teleported to Alice.

\subsection*{Entanglement fidelity of the network links}
\begin{figure*}[ht]
    \centering
    \includegraphics{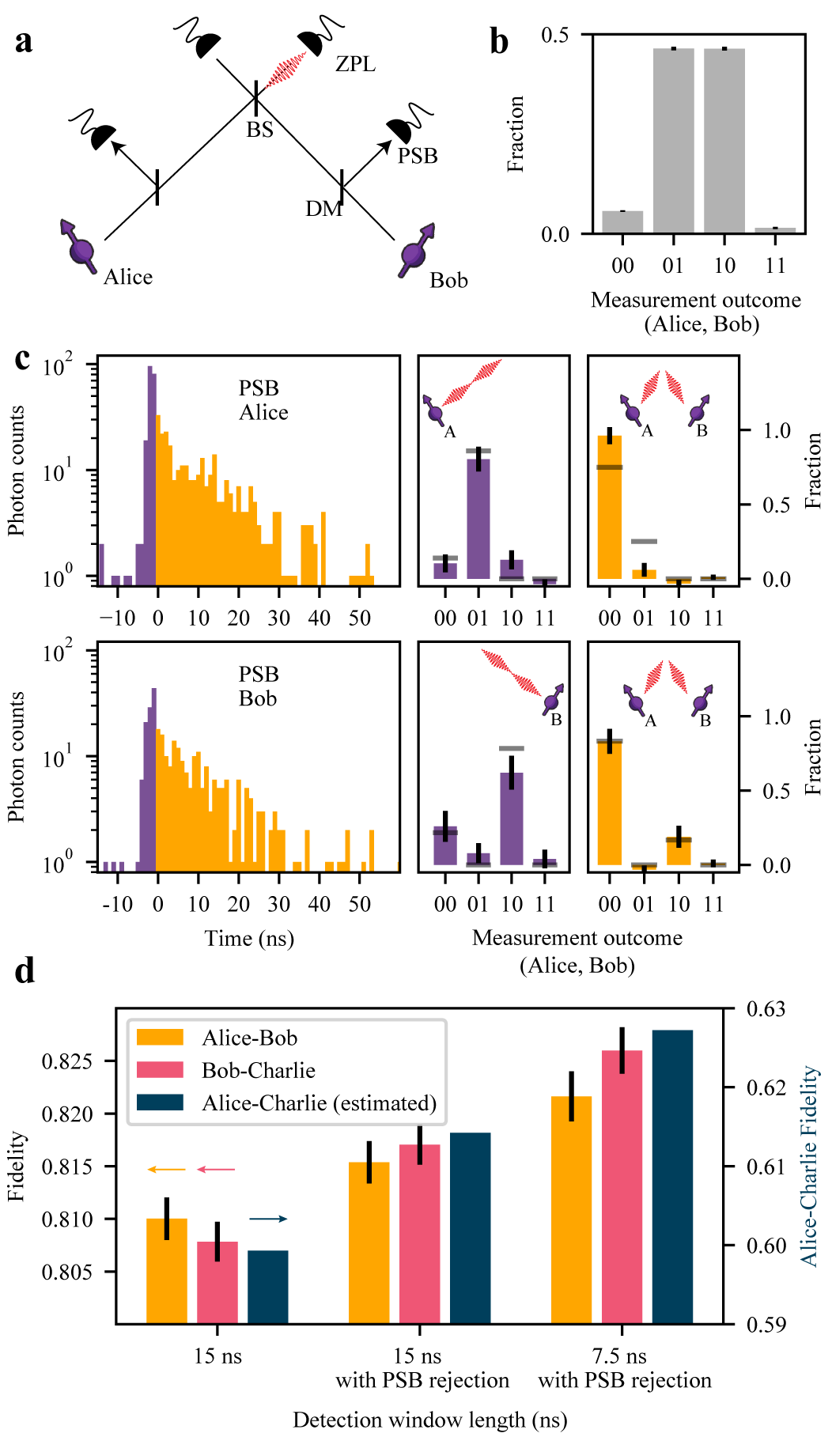}
    \caption{\textbf{High-fidelity entangled network links.} 
    \FigLet{a} Simplified schematic of the optical link used for generating entanglement between neighboring nodes. Photons emitted by the communication qubits are filtered by a dichroic mirror (DM) to separate the resonant (zero-phonon line, ZPL) photons (3\% of emission) from the off-resonant (phonon-side band, PSB) photons (97\% of emission). The resonant photons are sent to the beam splitter (BS); detection of a single photon at one of the ZPL detectors heralds successful generation of an entangled state between the two nodes.
    \FigLet{b} Measured correlations of the communication qubits in the computational basis, conditioned on a heralding event on the ZPL detectors.
    \FigLet{c} (left) Histograms of the PSB photon detection times on Alice (top) or Bob (bottom), conditioned on a simultaneous ZPL detection in the same entanglement generation attempt. Gray lines show expected correlations based on a quantum-optical model~\cite{teleport_suppl}.
    \FigLet{d} Measured fidelity of the network links, without PSB rejection (left), with PSB rejection (middle) and with PSB rejection plus shortened detection window (right). The dark blue bars indicate the corresponding expected fidelity on Alice-Charlie after entanglement swapping for each case ( ~\cite{teleport_suppl} ).}
    \label{fig:teleport_two}
\end{figure*}

A key parameter for quantum teleportation is the fidelity of the pre-shared entangled state between Alice and Charlie. As we generate this state by entanglement swapping, its fidelity is upper bounded by the errors on the individual links. Therefore, mitigating error sources on the individual links is critical. Our network generates entanglement between neighboring nodes using a single-photon protocol~\cite{cabrillo_creation_1999,bose_proposal_1999} in an optical-phase-stabilized architecture~\cite{pompili_realization_2021}. The building block of this protocol is a qubit-photon entangled state created at each node. To generate this entangled state we initialize the communication qubit in a superposition state $\ket{\psi} = \sqrt{\alpha}\ket{0} + \sqrt{1-\alpha}\ket{1}$ and apply a state-selective optical pulse that transfers the population from $\ket{0}$ to an optically excited state. Following spontaneous emission, the qubit state is entangled with photon number (0 or 1 photon). We perform this protocol on both nodes and interfere the resonant photonic states on a beam splitter (Figure \ref{fig:teleport_two}a). Detection of a single photon in one of the beam splitter output ports ideally heralds the generation of an entangled state $\ket{\psi} = (\ket{01} \pm \ket{10})/\sqrt{2}$, where the $\pm$ phase is set by which detector clicked. Figure \ref{fig:teleport_two}b displays the joint outcomes of qubit measurements in the computational basis after entanglement is heralded, showing the expected correlations.

The infidelity of the generated state has three main contributions: double $\ket{0}$ state occupancy, double optical excitation and finite distinguishability of the photons~\cite{humphreys_deterministic_2018,pompili_realization_2021, teleport_suppl}. In the case of double $\ket{0}$ state occupancy (which occurs with probability $\alpha$), both communication qubits are in the $\ket{0}$ state and have emitted a photon. Detection of one of these photons leads to false heralding of an entangled state. The second effect, double excitation, is due to the finite length of the optical pulse compared to the emitter’s optical lifetime. There is a finite chance that the communication qubit emits a photon during this pulse, is subsequently re-excited during the remainder of the pulse and then emits another photon resulting in the qubit state being entangled with two photons. Detection or loss of the first photon destroys the coherence of the qubit-photon entangled state and detection of the second photon can then falsely herald the generation of an entangled state.

Crucially, false heralding events due to double $\ket{0}$ state occupancy and double excitation are both accompanied by an extra emitted photon. Therefore, detection of this additional photon allows for unambiguous identification of such events and thus for real-time rejection of the corresponding false heralding signals. We implement this rejection scheme by monitoring the off-resonant phonon-side band (PSB) detection path on both setups during and after the optical excitation (see Figure \ref{fig:teleport_two}a). 

To investigate the effect of this scheme, we generate entanglement on the individual links and extract the entanglement heralding events for which the PSB monitoring flagged the presence of an additional photon. For these events, we again analyze the corresponding qubit measurements in the computational basis (Figure \ref{fig:teleport_two}c). 

We identify two separate regimes: one during the optical pulse (purple) and one after the optical pulse (yellow). When a photon is detected on Alice’s (Bob’s) PSB detector during the optical pulse we see that the outcome 01 (10) is most probable (purple data in Figure \ref{fig:teleport_two}c) showing that only one setup was in the $\ket 0$ state and thus that both detected photons originated from Alice (Bob). The detection of PSB photons during the optical pulse thus primarily flags double excitation errors. In contrast, when a photon is detected after the optical pulse in either Alice’s or Bob’s PSB detector, the outcome 00 is most probable (yellow data in Figure \ref{fig:teleport_two}c), indicating that both setups were in the $\ket{0}$ state and both emitted one photon. PSB photon detection after the optical pulse thus flags the double $\ket{0}$ state occupancy error. We find similar results to Figure \ref{fig:teleport_two}c for the entangled states generated on the Bob-Charlie link, see~\cite{teleport_suppl}.
The improvement in fidelity from rejecting these false heralding events in our experiment is set by the combined probability of occurrence ($\approx$ 9 \%, see~\cite{teleport_suppl}) multiplied by the probability to flag them (given here by the total PSB photon detection efficiency of $\approx$ 10\%).

The third main source of infidelity, the finite distinguishability, can arise from frequency detunings between the emitted photons~\cite{legero_time-resolved_2003}. While most of these detunings are eliminated upfront by the charge-resonance (CR) check before the start of the protocol~\cite{teleport_suppl}, the communication qubits may still be subject to a small amount of spectral diffusion. In our single-photon protocol, this leads to dephasing that is stronger for photons that are detected later relative to the optical pulse. By shortening our detection window, we can increase the fidelity of the entangled state at the expense of a lower entangling rate. For the experiments below (unless mentioned differently) we use a detection window length of 15 ns.
Figure \ref{fig:teleport_two}d summarizes the measured improvements on the individual links. For the teleporter, we estimate that their combined effect is an increase in Alice-Charlie entangled state fidelity by $\approx$ 3\%. This increase is instrumental in pushing the teleportation fidelity above the classical bound.

\subsection*{Memory qubit coherence}
In the preparation of the teleporter it is crucial that the first entangled state between Alice and Bob is reliably preserved on the memory qubit while the second link between Bob and Charlie is being generated. For this reason we abort the sequence and start over when the second entangled state is not heralded within a fixed number of attempts, the timeout.

 The \Cteen $ $ memory qubits can be controlled with high fidelity via the communication qubit while they can be efficiently decoupled when no interaction is desired. Recent work showed that in a magnetic field of 189mT entanglement generation attempts with the communication qubit do not limit the memory dephasing time $T_2^\star$~\cite{pompili_realization_2021}, opening the door to significantly extending the memory preservation time with active coherence protection from the spin bath~\cite{bradley_ten-qubit_2019}. We realize this protection by integrating a decoupling $\pi$-pulse on the memory qubit into the experimental sequence that follows a heralding event, while ensuring that all phases that are picked up due to the probabilistic nature of the remote entangling process are compensated in real time (Figure \ref{fig:teleport_three}a).

In Figure \ref{fig:teleport_three}b we check the performance of this sequence by storing a superposition state on the memory qubit and measuring the Bloch vector length. We compare the results for the sequence with and without the decoupling $\pi$-pulse, and with and without entanglement attempts. We observe that without the decoupling pulse the decay of the Bloch vector length is not altered by the entanglement attempts, in line with previous findings~\cite{pompili_realization_2021}. In contrast, when we apply the decoupling pulse the decay is slowed down by more than a factor of 6, yielding a $N_{1/e}$ decay constant of $\approx$ 5300 entanglement attempts, the highest number reported to date for diamond devices. In addition, we observe a difference in the shape of the decay between the cases with and without entangling attempts, indicating that intrinsic decoherence is no longer the only limiting error source. The improved memory coherence enables us to use a timeout of 1000 entangling attempts, more than double that of Ref.~\cite{pompili_realization_2021}, which doubles the entanglement swapping rate.

\subsection*{Memory qubit readout}
High-fidelity memory qubit readout is required both in the preparation of the teleporter (at Bob) and during the teleportation protocol itself (at Charlie). The memory qubit is read out by mapping its state onto the communication qubit using quantum logic followed by single-shot readout of the communication qubit using state-dependent optical excitation and detection ~\cite{cramer_repeated_2016}. Due to limited photon collection efficiency ($\approx$ 10\%) and finite cyclicity of the optical transition ($\approx$ 99\%), the communication qubit readout fidelity is different for $\ket{0}$ and $\ket{1}$. As a result, for random initial states the probability that the correct state was assigned is significantly larger if one or more photons were detected (assigned outcome 0) than if no photons were detected (assigned outcome 1) ~\cite{robledo_high-fidelity_2011}. In previous work we circumvented this issue by conditioning on obtaining the outcome 0 ~\cite{pompili_realization_2021}. However, this approach scales unfavorably, as it forces the protocol to prematurely abort with probability >50\% at each memory qubit readout. Therefore, to access more complex protocols with multiple memory qubit readouts, near-deterministic readout schemes are required.

\begin{figure*}[ht]
    \centering
    \includegraphics{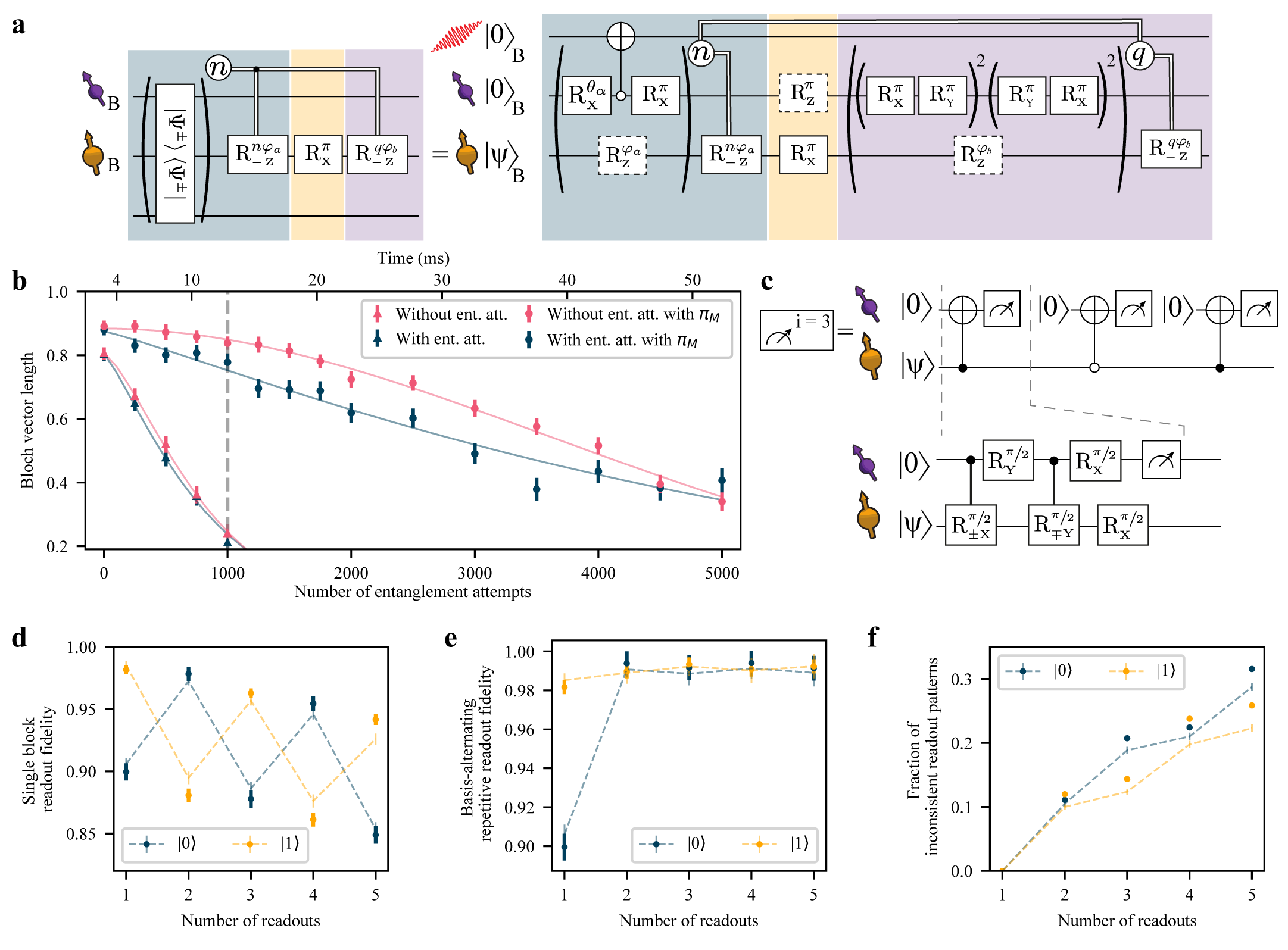}
    \caption{\textbf{Memory qubit coherence and readout.}
    \FigLet{a} Gate sequence on Bob for entanglement generation with the communication qubit while preserving states stored on the memory qubit. Entanglement generation attempts are repeated until success or a predetermined timeout. Upon success, a phase feed-forward is applied to maintain the correct reference frame of the memory qubit~\cite{pompili_realization_2021}, followed by a decoupling pulse on the memory qubit. The decoupling $\pi_\mathrm{M}$ pulse causes a Z-rotation on the communication qubit. Afterwards, we rephase the memory qubit for the same amount of time as it took to herald entanglement while applying an XY8 decoupling sequence on the communication qubit and we end with another phase feed-forward on the memory qubit to compensate for any phase picked up during this decoupling. 
    \FigLet{b} Bloch vector length of a superposition state stored on the memory qubit for different number of entanglement attempts or a time-equivalent wait element. In the case of no decoupling (no $\pi_\mathrm{M}$) on the memory qubit, the gates in the yellow shaded box in (a) are left out. The gray dashed line indicates the chosen timeout of 1000 entanglement attempts. 
    \FigLet{c} Gate sequence for the basis-alternating repetitive readout of the memory qubit. 
    \FigLet{d} Readout fidelity for each readout repetition, for state $\ket{0}$ and $\ket{1}$.
    \FigLet{e} Readout fidelity of the basis-alternating repetitive readout scheme for different number of readout repetitions. 
    \FigLet{f} Fraction of inconsistent readout patterns for different number of readout repetitions. In (d-f) the dashed lines show a numerical model using measured parameters.}
    \label{fig:teleport_three}
\end{figure*}

We resolve this challenge by introducing a basis-alternating repetitive readout for the memory qubit (see Figure \ref{fig:teleport_three}c). The key point of this readout strategy is, in contrast to earlier work ~\cite{jiang_repetitive_2009}, to alternatingly map the computational basis states of the memory qubit to the communication qubit state $\ket{0}$. Figure \ref{fig:teleport_three}d shows the readout fidelities of the $n$-th readout repetition for the two initial states for the memory qubit on Bob (for Charlie, see ~\cite{teleport_suppl}). We clearly observe the expected alternating pattern due to the asymmetry of the communication qubit readout fidelities. Importantly, the readout fidelity decays only by $\approx $1\% per readout, showing that the readout is mostly non-demolition and multiple readouts are possible without losing the state. We model the readout procedure using measured parameters (see ~\cite{teleport_suppl}) and plot the model’s predictions as dashed lines in Figure \ref{fig:teleport_three}d-f.

Next, we assign the state using the first readout and continue the sequence only when the consecutive readouts are consistent with the first readout. The subsequent readouts therefore add confidence to the assignment in the case of consistent outcomes, while cases of inconsistent outcomes (which have a higher chance of indicating an incorrect assignment) are filtered out. In Figure \ref{fig:teleport_three}e we plot the readout fidelity resulting from this strategy for up to five readouts, with the corresponding rejected fraction due to inconsistent outcomes plotted in Figure \ref{fig:teleport_three}f. We observe that using two readouts already eliminates most of the asymmetry, reducing the average infidelity from $\approx$ 6\% to below 1\%. At this point, the remaining observed infidelity mainly results from cases where the memory qubit was flipped during the first readout block. While adding further readout blocks does not lead to significant improvements in fidelity, each two additional readouts cut the amount of consistent outcomes by $\approx$ 10\%, due to the communication qubit readout infidelities and gate errors. For the experiments reported below (unless mentioned differently) we use two readout repetitions to benefit from a high average readout fidelity (Bob:99.2(4)\%, Charlie: 98.1(4)\%) and a high probability to continue the sequence (Bob and Charlie: $\approx$ 88\%).

\subsection*{Teleporting qubit states from Charlie to Alice}
\begin{figure*}[ht]
    \centering
    \includegraphics{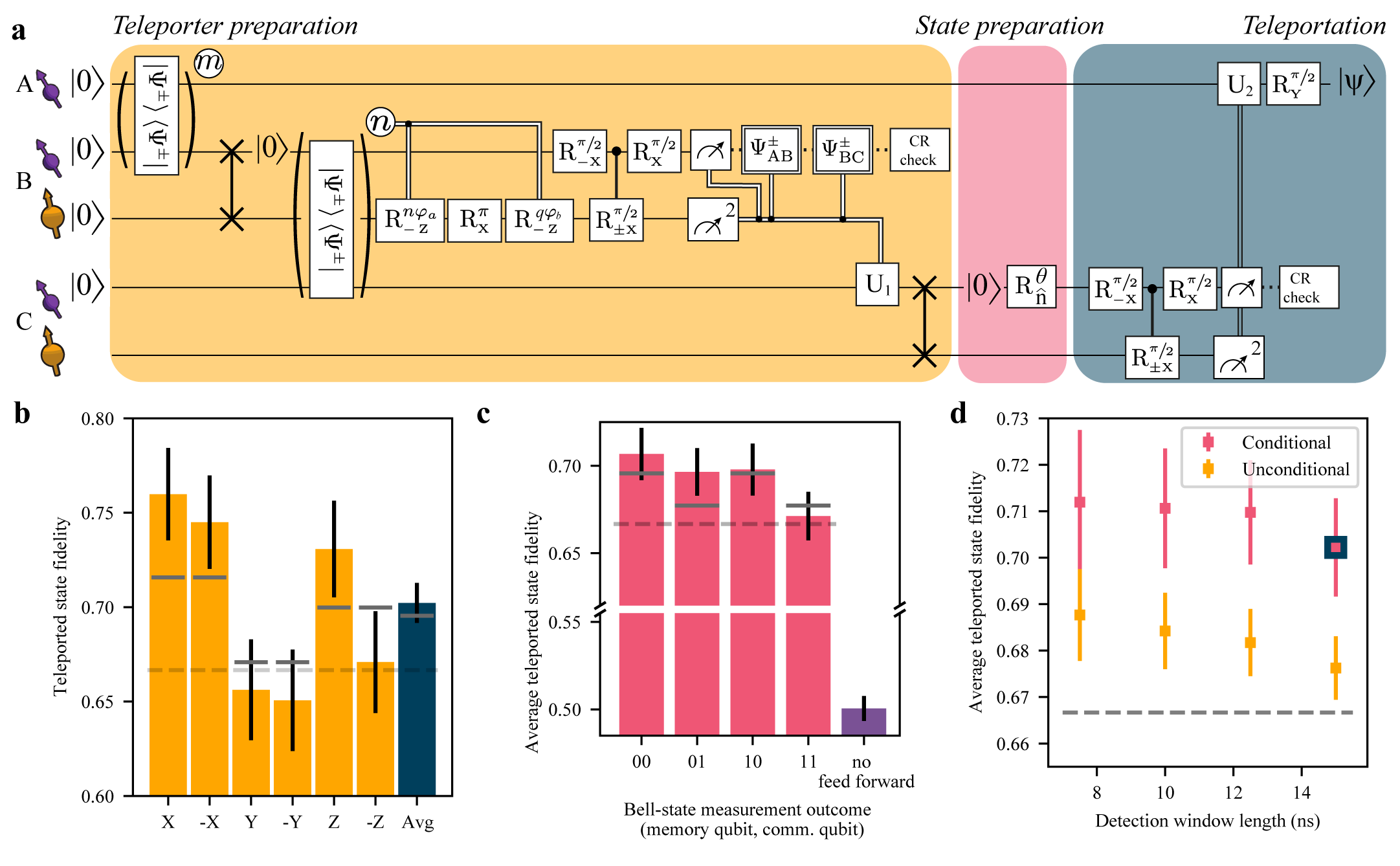}
    \caption{\textbf{Qubit teleportation between non-neighboring network nodes.} 
    \FigLet{a} Circuit diagram of the teleportation protocol using notation defined in Figure \ref{fig:teleport_three}. See ~\cite{teleport_suppl} for full circuit diagram. 
    \FigLet{b} Teleported state fidelities for the six cardinal states and their average. The gray lines show the expected fidelities from simulations. The dashed lines in (b-d) represents the classical bound of $2/3$. 
    \FigLet{c} Average teleported state fidelity for the different outcomes of the Bell-state measurement on Charlie. The right-most bar shows the resulting fidelity when no feed forward operation on Alice would be applied. 
    \FigLet{d} Average state fidelity for a conditional and unconditional teleportation, for different detection window lengths of the two-node entanglement generation processes. The blue bordered data point is the same point as shown in (b). }
    \label{fig:teleport_four}
\end{figure*}

With all innovations described above implemented, we perform the protocol as shown in Figure \ref{fig:teleport_four}a. First we generate entanglement between Alice and Bob and store Bob’s part of the entangled state on the memory qubit using a compiled SWAP operation. Second, we generate entanglement between Bob and Charlie, while preserving the first entangled state on the memory qubit with the pulse sequence as described in Figure \ref{fig:teleport_three}a. Next, we perform a Bell-state measurement on Bob followed by a CR check. We continue the sequence if the communication qubit readout yields outcome 0, the memory qubit readout gives a consistent outcome pattern and the CR check is passed. At Charlie, we perform a quantum gate that depends on the outcome of the Bell-state measurement and on which detectors clicked during the two-node entanglement generation. Next, we swap the entangled state to the memory qubit. At this point the teleporter is ready and Alice and Charlie share an entangled state with an estimated fidelity of 0.61. 

Subsequently, we generate the qubit state to be teleported, $\ket{\psi}$, on Charlie’s communication qubit and run the teleportation protocol. First, a Bell-state measurement is performed on the communication and memory qubits at Charlie. With the exception of unconditional teleportation (discussed below), we only continue the sequence when we obtain a 0 outcome on the communication qubit, when we have a consistent readout pattern on the memory qubit and when Charlie passes the CR check. The outcomes of the Bell-state measurement are sent to Alice and by applying the corresponding gate operation we obtain $\ket{\psi}$ on Alice’s side.

We teleport the six cardinal states $(\pm \mathrm{X}, \pm \mathrm{Y}, \pm \mathrm{Z})$, which form an unbiased set ~\cite{van_enk_experimental_2007}, and measure the fidelity of the teleported states to the ideally prepared state (Figure \ref{fig:teleport_four}b). We find an average teleported state fidelity of $F =  $0.702(11) at an experimental rate of $1/$(117s). This value exceeds the classical bound of $2/3$ by more than three standard deviations, thereby proving the quantum nature of the protocol. We note that this value provides a lower bound to the true teleportation fidelity, as the measured fidelity is lowered by errors in the preparation of the qubit states at Charlie (estimated to be 0.5\%, see~\cite{teleport_suppl}).

The differences in fidelity between the teleported states arise from an interplay of errors in different parts of the protocol that either affect all three axes (depolarizing errors) or only two axes (dephasing errors). These differences are qualitatively reproduced by our model (gray bars in Figure \ref{fig:teleport_four}b). 
In Figure \ref{fig:teleport_four}c we plot the teleportation fidelity for each possible outcome of the Bell-state measurement. Due to the basis-alternating repetitive readout, the dependence on the second bit (from the memory qubit readout) is small, whereas for the first bit (communication qubit readout) the best teleported state fidelity is achieved for outcome 0 due to the asymmetric readout fidelities. We also analyze the case in which no feed-forward is applied at Alice ~\cite{teleport_suppl}; as expected, the average state fidelity reduces to a value consistent with a fully mixed state (fidelity $F = $0.501(7)), emphasizing the critical role of the feed-forward in the teleportation protocol.

Finally, we demonstrate that the network can achieve unconditional teleportation between Alice and Charlie. Unconditional teleportation requires that, following preparation of the teleporter by establishing the remote entangled state, the protocol runs deterministically (each qubit state prepared at Charlie ends up at Alice) while surpassing the classical fidelity bound. We thus require that the Bell-state measurement at Charlie and the subsequent feed-forward operations are performed deterministically. To this end, we revise the protocol at Charlie to accept both communication qubit outcomes, use all memory qubit readout patterns including the inconsistent ones and disregard the outcome of the CR check after the Bell-state measurement. Using this fully deterministic Bell-state measurement lowers the average teleportation fidelity by a few percents (Figure \ref{fig:teleport_four}d). At the same time, shortening the detection windows of the two-node entanglement generation is expected to yield an improvement in the fidelity, as discussed above. We find indeed that the average unconditional teleportation fidelity increases with shorter window lengths, reaching $F =  $0.688(10) for a length of 7.5 ns and a rate of $1/$(100 s). The current quantum network is thus able to perform teleportation beyond the classical bound, even under the strict condition that every state inserted into the teleporter be transferred.

\subsection*{Outlook}
In this work we have realized unconditional qubit teleportation between non-neighboring nodes in a quantum network. The innovations introduced here on memory qubit readout and protection during entanglement generation, as well as the real-time rejection of false heralding signals, will be instrumental in exploring more complex protocols ~\cite{wehner_quantum_2018, Ben-Or2006,Arora2019,van_meter_quantum_2014,broadbent_universal_2009}. Also, these methods can be readily transferred to other platforms such as the group-IV color centers in diamond, the vacancy-related qubits in SiC and single rare-earth ions in solids ~\cite{rose_observation_2018, nguyen_quantum_2019, trusheim_transform-limited_2020, son_developing_2020,Lukin2020, kindem_control_2020,Chen2020}.

The development of an improved optical interface for the communication qubit~\cite{Ruf2021} will increase both the teleportation protocol rate and fidelity. Because of the improved memory qubit performance reported here, the network already operates close to the threshold where nodes can reliably deliver a remote entangled state while preserving previously stored quantum states in their memory qubits. With further improvements, for instance by integrating multi-pulse memory decoupling sequences~\cite{bradley_ten-qubit_2019} into the entanglement generation, demonstration of deterministic qubit teleportation may come within reach. In that case, the network is able to teleport a qubit state with unit efficiency at any given time, removing the need for heralding successful preparation of the teleporter and opening the door to exploring applications that call the teleportation routine multiple times.

Finally, by implementing a recently proposed link layer protocol ~\cite{dahlberg_link_2019}, qubit teleportation and applications making use of the teleportation primitive may be executed and tested on the network through platform-independent control software, an important prerequisite for a large-scale future network.


\subsection*{Acknowledgements}
We thank Stephanie Wehner, Tim Taminiau, Conor Bradley and Hugues de Riedmatten for discussions. We acknowledge financial support from the EU Flagship on Quantum Technologies through the project Quantum Internet Alliance (EU Horizon 2020, grant agreement no. 820445); from the Netherlands Organisation for Scientific Research (NWO) through a VICI grant (project no. 680-47-624) and the Zwaartekracht program Quantum Software Consortium (project no. 024.003.037/3368). S.B. acknowledges support from an Erwin-Schrödinger fellowship (QuantNet, no. J 4229-N27) of the Austrian National Science Foundation (FWF).

\bibliography{references}

\begin{thebibliography}{10}

\bibitem{bennett_teleporting_1993}
C.~H. Bennett, G.~Brassard, C.~Crépeau, R.~Jozsa, A.~Peres, and W.~K.
  Wootters, ``Teleporting an unknown quantum state via dual classical and
  {Einstein}-{Podolsky}-{Rosen} channels,'' {\em Physical Review Letters},
  vol.~70, pp.~1895--1899, Mar. 1993.

\bibitem{wehner_quantum_2018}
S.~Wehner, D.~Elkouss, and R.~Hanson, ``Quantum internet: {A} vision for the
  road ahead,'' {\em Science}, vol.~362, Oct. 2018.

\bibitem{Ben-Or2006}
M.~Ben-Or, C.~Cr{\'{e}}peau, D.~Gottesman, A.~Hassidim, and A.~Smith, ``{Secure
  multiparty quantum computation with (only) a strict honest majority},'' {\em
  Proceedings - Annual IEEE Symposium on Foundations of Computer Science,
  FOCS}, pp.~249--258, 2006.

\bibitem{Arora2019}
A.~S. Arora, J.~Roland, and S.~Weis, ``{Quantum weak coin flipping},'' {\em
  Proceedings of the Annual ACM Symposium on Theory of Computing},
  pp.~205--216, 2019.

\bibitem{bouwmeester_experimental_1997}
D.~Bouwmeester, J.-W. Pan, K.~Mattle, M.~Eibl, H.~Weinfurter, and A.~Zeilinger,
  ``Experimental quantum teleportation,'' {\em Nature}, vol.~390, pp.~575--579,
  Dec. 1997.

\bibitem{boschi_experimental_1998}
D.~Boschi, S.~Branca, F.~De~Martini, L.~Hardy, and S.~Popescu, ``Experimental
  {Realization} of {Teleporting} an {Unknown} {Pure} {Quantum} {State} via
  {Dual} {Classical} and {Einstein}-{Podolsky}-{Rosen} {Channels},'' {\em
  Physical Review Letters}, vol.~80, pp.~1121--1125, Feb. 1998.

\bibitem{furusawa_unconditional_1998}
A.~Furusawa, J.~L. Sørensen, S.~L. Braunstein, C.~A. Fuchs, H.~J. Kimble, and
  E.~S. Polzik, ``Unconditional {Quantum} {Teleportation},'' {\em Science},
  Oct. 1998.

\bibitem{olmschenk_quantum_2009}
S.~Olmschenk, D.~N. Matsukevich, P.~Maunz, D.~Hayes, L.-M. Duan, and C.~Monroe,
  ``Quantum {Teleportation} {Between} {Distant} {Matter} {Qubits},'' {\em
  Science}, Jan. 2009.

\bibitem{nolleke_efficient_2013}
C.~Nölleke, A.~Neuzner, A.~Reiserer, C.~Hahn, G.~Rempe, and S.~Ritter,
  ``Efficient {Teleportation} {Between} {Remote} {Single}-{Atom} {Quantum}
  {Memories},'' {\em Physical Review Letters}, vol.~110, p.~140403, Apr. 2013.

\bibitem{langenfeld_quantum_2021}
S.~Langenfeld, S.~Welte, L.~Hartung, S.~Daiss, P.~Thomas, O.~Morin,
  E.~Distante, and G.~Rempe, ``Quantum {Teleportation} between {Remote} {Qubit}
  {Memories} with {Only} a {Single} {Photon} as a {Resource},'' {\em Physical
  Review Letters}, vol.~126, p.~130502, Mar. 2021.

\bibitem{pfaff_unconditional_2014}
W.~Pfaff, B.~J. Hensen, H.~Bernien, S.~B.~v. Dam, M.~S. Blok, T.~H. Taminiau,
  M.~J. Tiggelman, R.~N. Schouten, M.~Markham, D.~J. Twitchen, and R.~Hanson,
  ``Unconditional quantum teleportation between distant solid-state quantum
  bits,'' {\em Science}, vol.~345, pp.~532--535, Aug. 2014.

\bibitem{bao_quantum_2012}
X.-H. Bao, X.-F. Xu, C.-M. Li, Z.-S. Yuan, C.-Y. Lu, and J.-W. Pan, ``Quantum
  teleportation between remote atomic-ensemble quantum memories,'' {\em
  Proceedings of the National Academy of Sciences}, vol.~109, pp.~20347--20351,
  Dec. 2012.

\bibitem{cabrillo_creation_1999}
C.~Cabrillo, J.~I. Cirac, P.~García-Fernández, and P.~Zoller, ``Creation of
  entangled states of distant atoms by interference,'' {\em Physical Review A},
  vol.~59, pp.~1025--1033, Feb. 1999.

\bibitem{bose_proposal_1999}
S.~Bose, P.~L. Knight, M.~B. Plenio, and V.~Vedral, ``Proposal for
  {Teleportation} of an {Atomic} {State} via {Cavity} {Decay},'' {\em Physical
  Review Letters}, vol.~83, pp.~5158--5161, Dec. 1999.

\bibitem{pompili_realization_2021}
M.~Pompili, S.~L.~N. Hermans, S.~Baier, H.~K.~C. Beukers, P.~C. Humphreys,
  R.~N. Schouten, R.~F.~L. Vermeulen, M.~J. Tiggelman, L.~d.~S. Martins,
  B.~Dirkse, S.~Wehner, and R.~Hanson, ``Realization of a multinode quantum
  network of remote solid-state qubits,'' {\em Science}, vol.~372,
  pp.~259--264, Apr. 2021.

\bibitem{humphreys_deterministic_2018}
P.~C. Humphreys, N.~Kalb, J.~P.~J. Morits, R.~N. Schouten, R.~F.~L. Vermeulen,
  D.~J. Twitchen, M.~Markham, and R.~Hanson, ``Deterministic delivery of remote
  entanglement on a quantum network,'' {\em Nature}, vol.~558, pp.~268--273,
  June 2018.

\bibitem{teleport_suppl}
``{Supplementary materials for `Qubit teleportation between non-neighboring
  nodes in a quantum network'}.''

\bibitem{legero_time-resolved_2003}
T.~Legero, T.~Wilk, A.~Kuhn, and G.~Rempe, ``Time-resolved two-photon quantum
  interference,'' {\em Applied Physics B}, vol.~77, pp.~797--802, Dec. 2003.

\bibitem{bradley_ten-qubit_2019}
C.~Bradley, J.~Randall, M.~Abobeih, R.~Berrevoets, M.~Degen, M.~Bakker,
  M.~Markham, D.~Twitchen, and T.~Taminiau, ``A {Ten}-{Qubit} {Solid}-{State}
  {Spin} {Register} with {Quantum} {Memory} up to {One} {Minute},'' {\em
  Physical Review X}, vol.~9, p.~031045, Sept. 2019.

\bibitem{cramer_repeated_2016}
J.~Cramer, N.~Kalb, M.~A. Rol, B.~Hensen, M.~S. Blok, M.~Markham, D.~J.
  Twitchen, R.~Hanson, and T.~H. Taminiau, ``Repeated quantum error correction
  on a continuously encoded qubit by real-time feedback,'' {\em Nature
  Communications}, vol.~7, p.~11526, May 2016.

\bibitem{robledo_high-fidelity_2011}
L.~Robledo, L.~Childress, H.~Bernien, B.~Hensen, P.~F.~A. Alkemade, and
  R.~Hanson, ``High-fidelity projective read-out of a solid-state spin quantum
  register,'' {\em Nature}, vol.~477, pp.~574--578, Sept. 2011.

\bibitem{jiang_repetitive_2009}
L.~Jiang, J.~S. Hodges, J.~R. Maze, P.~Maurer, J.~M. Taylor, D.~G. Cory, P.~R.
  Hemmer, R.~L. Walsworth, A.~Yacoby, A.~S. Zibrov, and M.~D. Lukin,
  ``Repetitive {Readout} of a {Single} {Electronic} {Spin} via {Quantum}
  {Logic} with {Nuclear} {Spin} {Ancillae},'' {\em Science}, vol.~326,
  pp.~267--272, Oct. 2009.

\bibitem{van_enk_experimental_2007}
S.~J. van Enk, N.~Lütkenhaus, and H.~J. Kimble, ``Experimental procedures for
  entanglement verification,'' {\em Physical Review A}, vol.~75, p.~052318, May
  2007.

\bibitem{van_meter_quantum_2014}
R.~Van~Meter, {\em Quantum networking}.
\newblock Networks and telecommunications series, John Wiley \& Sons, 2014.

\bibitem{broadbent_universal_2009}
A.~Broadbent, J.~Fitzsimons, and E.~Kashefi, ``Universal {Blind} {Quantum}
  {Computation},'' in {\em 2009 50th {Annual} {IEEE} {Symposium} on
  {Foundations} of {Computer} {Science}}, pp.~517--526, Oct. 2009.

\bibitem{rose_observation_2018}
B.~C. Rose, D.~Huang, Z.-H. Zhang, P.~Stevenson, A.~M. Tyryshkin,
  S.~Sangtawesin, S.~Srinivasan, L.~Loudin, M.~L. Markham, A.~M. Edmonds, D.~J.
  Twitchen, S.~A. Lyon, and N.~P.~d. Leon, ``Observation of an environmentally
  insensitive solid-state spin defect in diamond,'' {\em Science}, vol.~361,
  pp.~60--63, July 2018.

\bibitem{nguyen_quantum_2019}
C.~Nguyen, D.~Sukachev, M.~Bhaskar, B.~Machielse, D.~Levonian, E.~Knall,
  P.~Stroganov, R.~Riedinger, H.~Park, M.~Lončar, and M.~Lukin, ``Quantum
  {Network} {Nodes} {Based} on {Diamond} {Qubits} with an {Efficient}
  {Nanophotonic} {Interface},'' {\em Physical Review Letters}, vol.~123,
  p.~183602, Oct. 2019.

\bibitem{trusheim_transform-limited_2020}
M.~E. Trusheim, B.~Pingault, N.~H. Wan, M.~Gündoğan, L.~De~Santis,
  R.~Debroux, D.~Gangloff, C.~Purser, K.~C. Chen, M.~Walsh, J.~J. Rose, J.~N.
  Becker, B.~Lienhard, E.~Bersin, I.~Paradeisanos, G.~Wang, D.~Lyzwa, A.~R.-P.
  Montblanch, G.~Malladi, H.~Bakhru, A.~C. Ferrari, I.~A. Walmsley,
  M.~Atatüre, and D.~Englund, ``Transform-{Limited} {Photons} {From} a
  {Coherent} {Tin}-{Vacancy} {Spin} in {Diamond},'' {\em Physical Review
  Letters}, vol.~124, p.~023602, Jan. 2020.

\bibitem{son_developing_2020}
N.~T. Son, C.~P. Anderson, A.~Bourassa, K.~C. Miao, C.~Babin, M.~Widmann,
  M.~Niethammer, J.~Ul~Hassan, N.~Morioka, I.~G. Ivanov, F.~Kaiser,
  J.~Wrachtrup, and D.~D. Awschalom, ``Developing silicon carbide for quantum
  spintronics,'' {\em Applied Physics Letters}, vol.~116, p.~190501, May 2020.

\bibitem{Lukin2020}
D.~M. Lukin, M.~A. Guidry, and J.~Vu{\v{c}}kovi{\'{c}}, ``{Integrated Quantum
  Photonics with Silicon Carbide: Challenges and Prospects},'' {\em PRX
  Quantum}, vol.~1, no.~2, pp.~1--19, 2020.

\bibitem{kindem_control_2020}
J.~M. Kindem, A.~Ruskuc, J.~G. Bartholomew, J.~Rochman, Y.~Q. Huan, and
  A.~Faraon, ``Control and single-shot readout of an ion embedded in a
  nanophotonic cavity,'' {\em Nature}, vol.~580, pp.~201--204, Apr. 2020.

\bibitem{Chen2020}
S.~Chen, M.~Raha, C.~M. Phenicie, S.~Ourari, and J.~D. Thompson, ``{Parallel
  single-shot measurement and coherent control of solid-state spins below the
  diffraction limit},'' {\em Science}, vol.~370, no.~6516, pp.~592--595, 2020.

\bibitem{Ruf2021}
M.~Ruf, N.~H. Wan, H.~Choi, D.~Englund, and R.~Hanson, ``{Quantum networks
  based on color centers in diamond},'' {\em Journal of Applied Physics},
  vol.~130, no.~7, p.~070901, 2021.

\bibitem{dahlberg_link_2019}
A.~Dahlberg, M.~Skrzypczyk, T.~Coopmans, L.~Wubben, F.~Rozp\c{e}dek,
  M.~Pompili, A.~Stolk, P.~Pawe\l{}czak, R.~Knegjens, J.~de~Oliveira~Filho,
  R.~Hanson, and S.~Wehner, ``A link layer protocol for quantum networks,'' in
  {\em Proceedings of the {ACM} {Special} {Interest} {Group} on {Data}
  {Communication}}, {SIGCOMM} '19, (New York, NY, USA), pp.~159--173,
  Association for Computing Machinery, Aug. 2019.

\bibitem{Note1}


\bibitem{Note2}


\bibitem{kalb_dephasing_2018}
N.~Kalb, P.~C. Humphreys, J.~J. Slim, and R.~Hanson, ``Dephasing mechanisms of
  diamond-based nuclear-spin memories for quantum networks,'' {\em Physical
  Review A}, vol.~97, p.~062330, June 2018.

\end{thebibliography}
\bibliographystyle{my_plain}


\setcounter{equation}{0}
\setcounter{figure}{0}

\onecolumngrid
\fontsize{12pt}{10pt}

\centerline{\Large Supplementary Materials for}
\centerline{ \LARGE Qubit teleportation between non-neighboring nodes in a quantum network}
\vspace{1em}
\centerline{S.L.N.Hermans$^\ast$, M.Pompili$^\ast$, H.K.C.Beukers, S.Baier, J.Borregaard \& R.Hanson$^\dagger$ }
\centerline{QuTech and Kavli Institute of Nanoscience, Delft University of Technology}
\centerline{2628 CJ Delft, The Netherlands}
\vspace{1em}
\centerline{$^\ast$These authors contributed equally to this work}
\centerline{$^\dagger$Corresponding author}

\section{Full gate circuit}
\begin{sidewaysfigure}
	\centering
	\includegraphics[width=\linewidth]{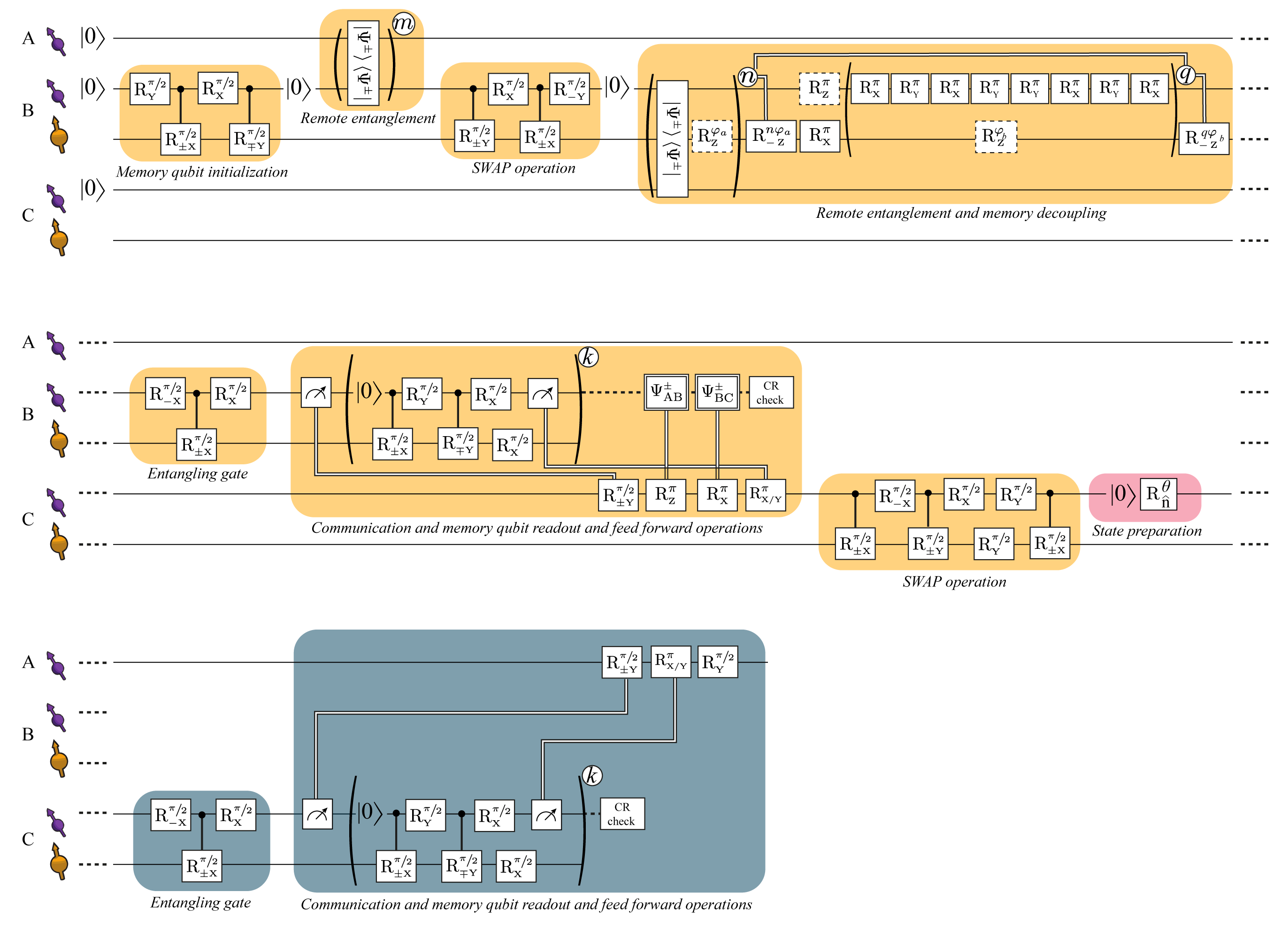}
	\caption{\label{fig:supp_full_gate_circuit} Full gate circuit for the teleportation protocol, see text for the description of each element.}
\end{sidewaysfigure}

Our quantum network consists of three nodes, Alice, Bob and Charlie. In the experiment, we will teleport a qubit from Charlie to Alice, two non-neighboring nodes. The full gate circuit is shown in Figure \ref{fig:supp_full_gate_circuit}. Prior to the sequence, we do a Charge-Resonance (CR) check on each node to ensure that the communication qubits are in the correct charge state (NV$^-$) and on resonance with the control lasers. Once all the nodes have passed this check, we do a first round of optical phase stabilization of the interferometers, which enables the entanglement generation using the single click protocol~\cite{cabrillo_creation_1999, bose_proposal_1999, humphreys_deterministic_2018, pompili_realization_2021}. After these preparation steps, the sequence is triggered on all setups. 
 
On Bob, we initialize the memory qubit into $\ket{0}$ using the communication qubit~\cite{cramer_repeated_2016}. Next, we generate entanglement between the communication qubits of Alice and Bob. When entanglement is heralded, we perform a SWAP operation to store Bob's part of the entangled state on the memory qubit. 
 
We continue with a second round of phase stabilization (not shown in the circuit) and generate entanglement between the communication qubits of Bob and Charlie. Each entanglement attempt slightly decoheres the memory qubit, therefore we limit the number of attempts by a timeout. If we do not succeed within the timeout, we abort the sequence and start over. 
 
During entanglement generation, the memory qubit of Bob picks up an average phase $n \varphi_{a}$ dependent on the number of entanglement attempts $n$. Due to the probabilistic nature of the entanglement generation process, we do not know which attempt will be successful, therefore this phase is unknown at the start of the sequence. To maintain the correct reference frame of the memory qubit this phase needs to be corrected in real-time before any other gate can be applied to the memory qubit. We perform this real-time correction by changing the time between pulses on the communication qubit~\cite{pompili_realization_2021}. After the phase correction, the decoupling pulse is applied to the memory qubit via the communication qubit. The back-action of this gate causes a Z-rotation on the communication qubit. To rephase the memory qubit, we wait for the same amount of time as it took to herald the second entangled state while decoupling the communication qubit. This imprints a phase $q \varphi_{b}$ on the memory qubit, which we compensate in an analogous way.  
 
Bob now shares two entangled states; his memory qubit is entangled with Alice and his communication qubit with Charlie. To establish an entangled state between Alice and Charlie we perform a Bell-state measurement on the two qubits of Bob. To do so, we entangle the communication and memory qubits and do a measurement on the communication qubit. We map its state onto the communication qubit and measure the communication qubit. In the basis-alternating repetitive readout, we repeat the measurement sequence twice. During the first readout we map the $\ket{0}$ state to the $\ket{0}$ state of the communication qubit, and in the second readout we map $\ket{1}$ to $\ket{0}$. The first outcome is used to assign the state and the second outcomes serves as a check. By continuing the sequence only when we measure consistent patterns (for instance $(m1, m2) = (1,0)$) we increase our average readout fidelity. After the readout procedure, we perform a CR check on Bob to filter out any event where Bob was in the wrong charge state. 
 
Bob communicates to Charlie which gate operation should be done to obtain the correct entangled state. Which operation is required is determined by the outcomes of the Bell-state measurement on Bob and by which detector heralded the individual links. Charlie performs the feed-forward gate operation and subsequently stores its part of the entangled state on the memory qubit using a SWAP gate. At this point in the sequence the teleporter is ready.
 
To prepare the state that is to be teleported, we initialize the communication qubit at Charlie and perform the desired qubit rotation.
 
To teleport the qubit, we perform a Bell-state measurement on the qubits of Charlie. Locally, we entangle the communication qubit with the memory qubit. We readout the communication qubit and use the basis-alternating repetitive readout for the memory qubit. Additionally, we do a CR check on Charlie. Charlie communicates the results of the Bell-state measurement to Alice, and Alice performs a feed-forward operation to obtain the teleported state. 
 
To verify the teleported state, we measure the state of Alice in the corresponding basis. To prevent any bias in the tomography we measure in both directions, e.g. when we teleport $\ket{+Z}$ we measure both along +Z and -Z axes. 
 
\section{Experimental setup}

The basics of the experimental setup are described in~\cite{pompili_realization_2021}. In the current experiment, Charlie has access to a carbon-13 nuclear spin that acts as a memory qubit. The parameters used for the memory qubits of Bob and Charlie can be found in Table~\ref{tab:c13_couplings}. Additionally, we have set up a classical communication channel between Charlie and Alice such that Charlie can directly send the results of the Bell-state measurement to Alice. 

\section{Tailored heralding of the remote entangled states}

In the main text we describe several noise mechanisms that reduce the remote two-node entangled state fidelity. Two of these noise mechanisms, double $\ket{0}$ occupancy and double optical excitation, are accompanied by the emission of an extra photon. This extra photon can be detected using the local phonon-side band (PSB) detectors. By monitoring the PSB detectors, we can real-time reject false heralding events. 

\begin{figure}[h]
	\centering
	\includegraphics[width=0.8\linewidth]{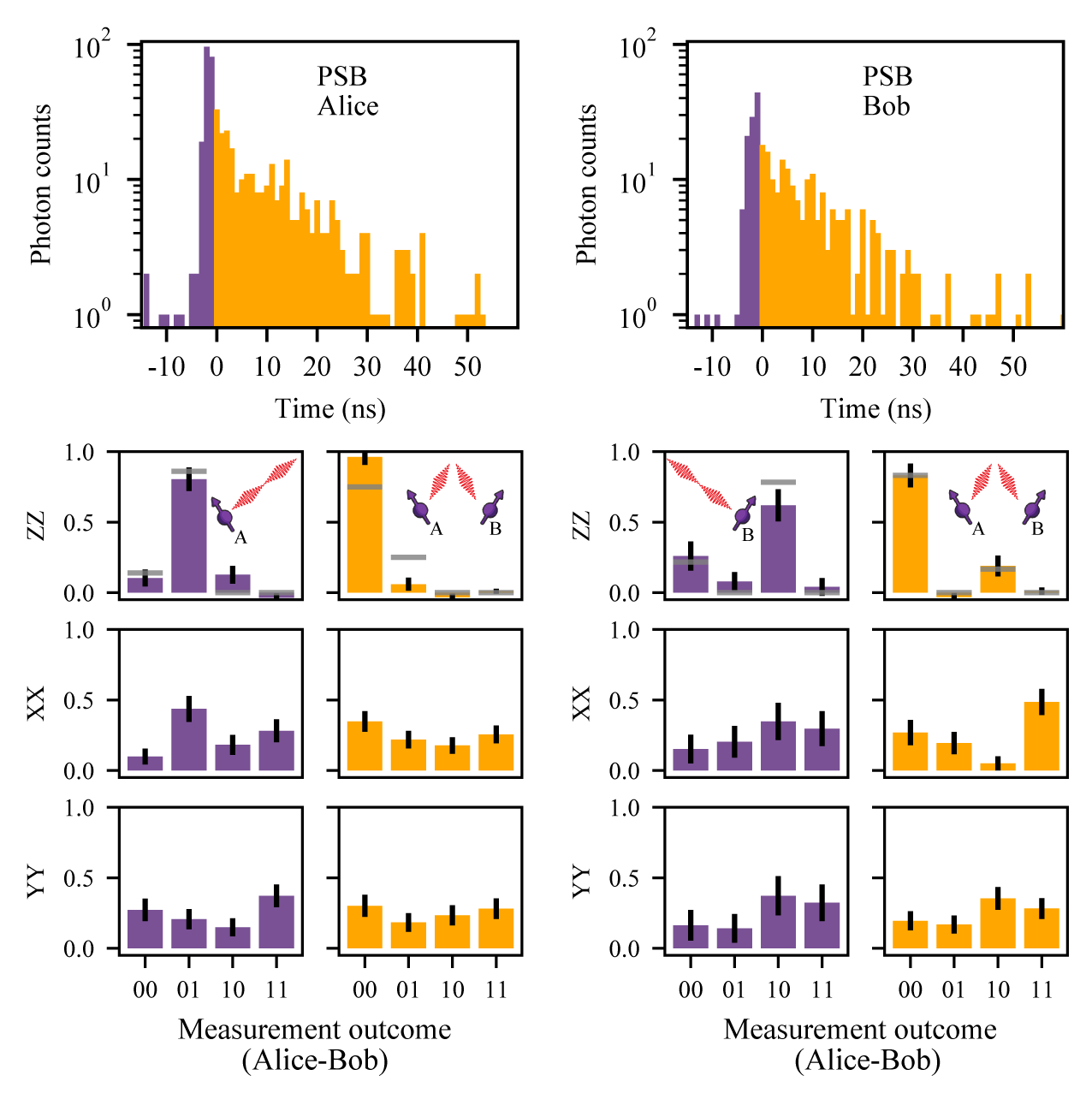}
	\caption{\label{fig:supp_psb_filter_AB} (Top) Histograms of the detected PSB photons conditioned on a simultaneous ZPL detection in the entanglement generation attempt, for Alice (left) and Bob (right). (Bottom) Corresponding measured correlations in all bases. The gray bars in the Z basis represent the simulated values. For the X and Y bases, one would expect a probability of 0.25 for all outcomes.}
\end{figure}

\begin{figure}[h]
	\centering
	\includegraphics[width=0.8\linewidth]{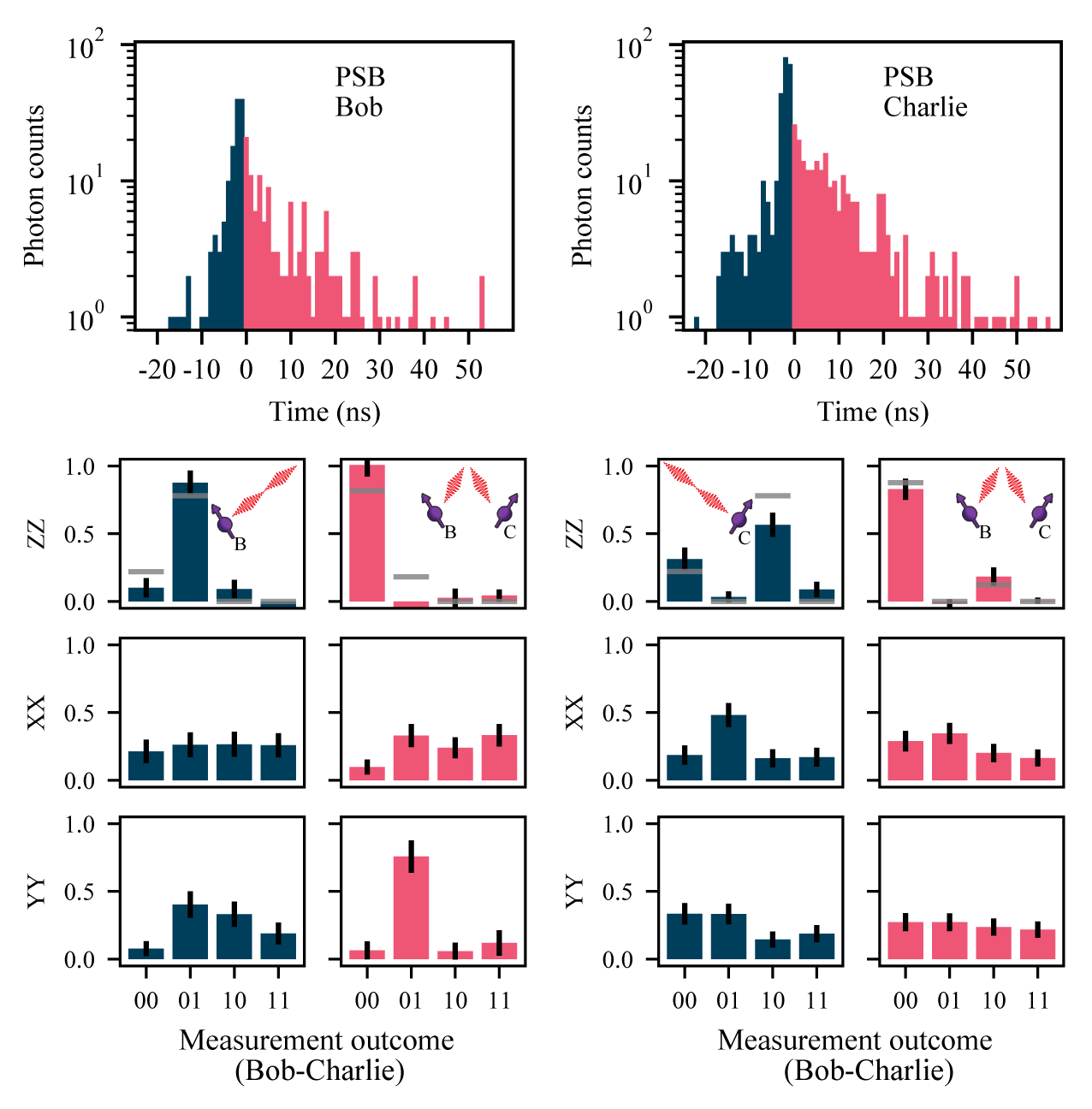}
	\caption{\label{fig:supp_psb_filter_BC} (Top) Histograms of the detected PSB photons conditioned on a simultaneous ZPL detection in the entanglement generation attempt, for Bob (left) and Charlie (right). (Bottom) Corresponding measured correlations in all bases. The gray bars in the Z basis represent the simulated values. For the X and Y basis, one would expect a probability of 0.25 for all outcomes.}
\end{figure}

In figures~\ref{fig:supp_psb_filter_AB} and~\ref{fig:supp_psb_filter_BC}, we plot the histograms of the detection times of the PSB photons conditioned on a simultaneous heralding (zero-phonon line, ZPL) photon detection in the same entanglement generation attempt, for the Alice-Bob and Bob-Charlie entangled link respectively. The correlations are measured in the computational (or Z) basis, and in the X and Y basis. In the computational basis we see the behavior dependent on the detection time of the PSB photon as described in the main text together with the simulations (gray bars). In the X and Y basis, all outcomes are equally probable, and the quantum correlations are washed out. 

From the data collected, we can extract the probability to detect these additionally emitted PSB photons. We assume the dark counts of the detectors to be negligible, the PSB detections during the pulse to be fully dominated by the double optical excitation error, and the PSB detections after the pulse to be only caused by double $\ket{0}$ occupancy. By correcting for the PSB detection efficiency, we can estimate the probability for double $\ket{0}$ occupancy and double optical excitation errors. The results are given in Table~\ref{tab:alpha_pde}. The double $\ket{0}$ state error is expected to occur with probability $\alpha$. The extracted numbers correspond well to the parameter values we use during remote entanglement generation ($\alpha_{\text{Alice}} = 0.07, \alpha_{\text{Bob}} = 0.05, \alpha_{\text{Charlie}} = 0.10$). The probability for the double optical excitation to occur depends on the shape and the amplitude of the optical excitation pulse, and differs per node. 

\subsection{Numerical model}
We compare our PSB detection data (previous section) to a numerical model. We model the NV center as a three level system with two stable ground states $\ket{0},\ket{1}$ and one excited state $\ket{e}$. The optical $\ket{0}\leftrightarrow\ket{e}$ transition is driven by a resonant laser pulse and is assumed to be a closed transition. The Hamiltonian describing the dynamics of the system in a suitable rotating frame is 
\begin{equation} \label{eq:hamil1}
\hat{H}=\Omega(t)\ket{e}\bra{0}+\Omega^*(t)\ket{0}\bra{e},
\end{equation}
where $\Omega(t)$ describes the (time-dependent) driving of the optical transition. From the excited state, the NV can spontaneously emit a photon and decay to $\ket{0}$. Without specifying the particular mode this photon is emitted in, we simply model such an emission with a Lindblad jump operator of the form $\hat{L}_1=\sqrt{\gamma}\ket{0,1_{p}}\bra{e}$. Here $\gamma$ is the rate of spontaneous emission, $\ket{0,1_{p}}$ denotes the state where the NV is in state $\ket{0}$ and one photon was emitted, and we use the convention that when not explicitly stated, there is no emitted photon i.e. $\ket{e}$ denotes the NV in state $\ket{e}$ with zero emitted photons. 

To account for double emission errors in the entanglement scheme, we expand the model by letting states $\ket{0,1_p},\ket{e,1_p}$ be coupled by a similar Hamiltonian as in Eq.~(\ref{eq:hamil1}) with the same coupling $\Omega(t)$. Double emission is then captured by a Lindblad jump operator $\hat{L}_2=\sqrt{\gamma}\ket{0,2_{p}}\bra{1_p,e}$. For the specific excitation pulses used in the experiment, we can then numerically solve the Master equation of the system in a basis of $\{\ket{0},\ket{e},\ket{0,1_p},\ket{e,1_p},\ket{0,2_p}\}$ to obtain the probability of zero ($P_0$), one ($P_1$), or two ($P_2$) photons being emitted from the system ($P_0+P_1+P_2=1$). Note that in this model, we neglect the probability of emitting more than two photons from the NV.   

Assuming an initial state $\sqrt{\alpha}\ket{0}+\sqrt{1-\alpha}\ket{1}$ of the NV center, the state after the optical excitation is then modeled as 
\begin{equation} 
\ket{\psi}=\sqrt{\alpha}\left(\sqrt{P_0}\ket{0}+\sqrt{P_1}\ket{0,1_p}+\sqrt{P_2}\ket{0,2_p}\right)+\sqrt{1-\alpha}\ket{1}. 
\end{equation}
The emitted photons are either PSB ($= 97\%$) or ZPL ($= 3\%)$ photons. We model this by performing a standard beam splitter transformation on the photonic modes. Letting $\hat{a}^{\dagger}$ be the creation operator of a photon ($\ket{1_p}=\hat{a}^{\dagger}\ket{0_p}$), we make the transformation $\hat{a}^{\dagger}\to\sqrt{P_{z}}\hat{a}^{\dagger}_{z}+\sqrt{1-P_{z}}\hat{a}^{\dagger}_{b}$, where $\hat{a}^{\dagger}_{z}$ ($\hat{a}^{\dagger}_{b}$) is the creation operator of a ZPL (PSB) photon and $P_{z} = 3\%$ . Consequently, $\ket{1_p}\to\sqrt{P_{z}}\ket{1_{z}}+\sqrt{1-P_{z}}\ket{1_{b}}$, where $\ket{1_{z}}$ ($\ket{1_{b}}$) is an emitted ZPL (PSB) photon. 

The photons can be emitted either inside or outside the detection time window, i.e. the time interval in which detected photons are accepted. This time interval is in general different for the PSB and ZPL photons. This results in the following transformations: 
\begin{eqnarray}
\ket{1_z}&\to&\sqrt{P_{dz,1}}\ket{1_{d,z}}+\sqrt{1-P_{dz,1}}\ket{1_{nd,z}} \\
\ket{1_b}&\to&\sqrt{P_{db,1}}\ket{1_{d,b}}+\sqrt{1-P_{db,1}}\ket{1_{nd,b}} \\
\ket{2_z}&\to&\sqrt{P_{dz,2}}\ket{2_{d,z}}+\sqrt{P_{dz,3}}\ket{1_{d,z}}\ket{1_{nd,z}}+\sqrt{1-P_{dz,2}-P_{dz,3}}\ket{2_{nd,z}} \\
\ket{2_b}&\to&\sqrt{P_{db,2}}\ket{2_{d,b}}+\sqrt{P_{db,3}}\ket{1_{d,b}}\ket{1_{nd,b}}+\sqrt{1-P_{db,2}-P_{db,3}}\ket{2_{nd,b}} \\
\ket{1_z}\ket{1_b}&\to&\sqrt{P_{dzb,1}}\ket{1_{d,z}}\ket{1_{d,b}}+\sqrt{P_{dzb,2}}\ket{1_{nd,z}}\ket{1_{d,b}}\\ \nonumber
&&+\sqrt{P_{dzb,3}}\ket{1_{d,z}}\ket{1_{nd,b}}+\sqrt{1-P_{dz,2}-P_{dz,2}-P_{dz,3}}\ket{1_{nd,z}}\ket{1_{nd,b}}.  
\end{eqnarray}
The probabilities $P_{dz,1}, P_{db,1},\ldots$ are defined in table~\ref{tab:tableS1} and are found through the numerical simulation described above.  

Finally, we model transmission loss with standard beam splitter transformations acting on the photon modes emitted in the detection window. Letting $\hat{a}^{\dagger}_{d,z}$ ($\hat{a}^{\dagger}_{d,b}$) be the creation operator of a ZPL (PSB) photon emitted in the detection time window, we make the transformations
\begin{eqnarray}
\hat{a}^{\dagger}_{\text{d,z}}\to\sqrt{\eta_{z}}\hat{a}^{\dagger}_{\text{d,z}}+\sqrt{1-\eta_{z}}\hat{a}^{\dagger}_{\text{nd,z}} \\
\hat{a}^{\dagger}_{\text{d,z}}\to\sqrt{\eta_{b}}\hat{a}^{\dagger}_{\text{d,b}}+\sqrt{1-\eta_{z}}\hat{a}^{\dagger}_{\text{nd,b}}.
\end{eqnarray}
where $\eta_{z}$ is the total transmission efficiency from the NV to the central beam splitter while $\eta_{b}$ is the total transmission and detection efficiency of the PSB photons. The operators $\hat{a}^{\dagger}_{\text{nd,z}}$ and $\hat{a}^{\dagger}_{\text{nd,b}}$ describe the lost/undetected modes. Tracing over the undetected modes, the output state of a single NV can be written as
\begin{equation} \label{eq:output}
\rho_{\psi}=\rho_0\otimes\ket{0}\bra{0}_\text{d,b}+\rho_1\otimes\ket{1}\bra{1}_\text{d,b}+\rho_2\otimes\ket{2}\bra{2}_\text{d,b},
\end{equation}   
where we have neglected any coherence between the photonic PSB modes since these are accompanied by undetected non-radiative decay (phonon emission). The unnormalized density matrices $\rho_0,\rho_1$, and $\rho_2$ describe the state of the NV center communication qubit and the ZPL photons emitted in the time window of the ZPL detectors and transmitted to the central beam splitter. In the limit $\eta_z\ll1$, we can neglect terms of $\ket{2_{\text{d,z}}}$ and these density matrices will all be of the form 
\begin{equation} \label{eq:sum_states}
\rho_j=\sum^{4}_{i=1}\ket{\phi_{i,j}}\bra{\phi_{i,j}},
\end{equation} 
where $\ket{\phi_{i,j}}=(a_{i,j}\ket{1}+b_{i,j}\ket{0})\ket{0_{z}}+c_{i,j}\ket{0}\ket{1_{\text{d,z}}}$ and $j=0,1,2$. In Eq.~(\ref{eq:sum_states}) i refers to the different number of lost undetected photons
\begin{align*}
& i = 1, \text{zero photons being lost} \\
& i = 2, \text{one ZPL photon being lost} \\
& i = 3, \text{one PSB photon being lost} \\
& i = 4, \text{two photons being lost, either two ZPL, two PSB or one ZPL and one PSB}\\
\end{align*}
and j to the number of detected PSB photons. We note that all $a_{i,1}$ and $a_{i,2}$ will be zero since $\rho_1$ and $\rho_2$ are accompanied by PSB photons (see Eq.~(\ref{eq:output})) meaning that the NV was in state $\ket{0}$. Furthermore, the only non-zero term in $\rho_2$ will be $b_{i,2}$ since two PSB photons were emitted, meaning that no ZPL photon was emitted since we neglect higher order emissions.  

The only term in Eq.~(\ref{eq:output}) from which remote spin-spin entanglement between two NVs can be created is $\rho_0\otimes\ket{0}\bra{0}_\text{d,b}$ since this does not have any detected PSB photons. However, PSB and ZPL photons that were emitted but not detected will still decrease the entangled state fidelity. Such events are responsible for the contributions of $\ket{\phi_{2,0}},\ket{\phi_{3,0}}$ and $\ket{\phi_{4,0}}$ in $\rho_0$. The only term where no PSB photons were emitted and no ZPL photons were undetected is $\ket{\phi_{1,0}}=\sqrt{1-\alpha}\ket{1}\ket{0_{\text{zpl}}}+\sqrt{\alpha}\ket{0}(\sqrt{P_0}\ket{0_{\text{zpl}}}+\sqrt{P_1P_{\text{zpl}}P_{\text{d,zpl}}}\ket{1_{\text{d,zpl}}})$. 

The combined state from the two NV centers before the central beam splitter is $\rho_{\psi}\otimes\tilde{\rho}_{\psi}$, where $\tilde{\rho}_{\psi}$ (the state of the second NV) is of the same form as in Eq.\ref{eq:output} but including that parameters such as initial rotation ($\alpha$), driving strength ($\Omega$) and transmission efficiencies ($\eta_z,\eta_b$) can be different for the two centers. Furthermore, we include a phase difference between the two paths to the central beam splitter. The central beam splitter is modeled as a perfect 50:50 beam splitter and the finite detection efficiency of the output detectors is assumed to be equal and can be directly included in the transmission efficiencies ($\eta_z$) while dark counts are negligible in the experiment and not included. Finally, we include non-perfect visibility between the ZPL photons by reducing the coherence between the output modes of the beam splitter by a factor $v$. This visibility is estimated from experimental data and can e.g. originate from slightly off-resonant driving of the NV centers.
 
\section{Memory qubit coherence Bob}

\begin{figure}
	\centering
	\includegraphics[width=\linewidth]{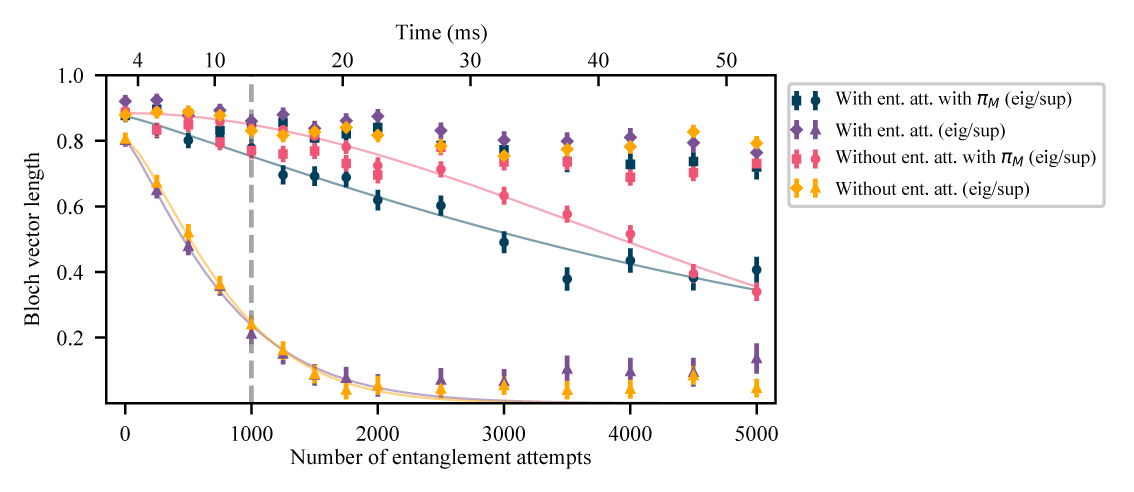}
	\caption{\label{fig:supp_mem_coh} Coherence of Bob's memory qubit for superposition states (triangles and circles) and eigenstates (squares and diamonds). We perform the sequence as described in the main text with and without the decoupling pulse $\pi_M$ on the memory qubit, the dark blue and purple points respectively. Additionally, we perform the sequence with a wait time instead of entanglement attempts with (pink points) and without the decoupling pulse (yellow points). The gray dashed line indicates the timeout of the entanglement generation process used in the teleportation protocol.}
\end{figure}

We use the sequence described in Figure 3a of the main text to preserve the state of the memory qubit during entanglement attempts. To characterize the decoupling sequence, we compare it to the sequence where we do not apply the decoupling pulse on the memory qubit and/or the sequence where we idle instead of performing entanglement attempts. We characterize the coherence of the memory qubit by storing the six cardinal states. We average the results for the eigenstates ($\ket{0}, \ket{1}$) and superposition states ($\ket{\pm X}$ and  $\ket{\pm Y}$). In Figure \ref{fig:supp_mem_coh} we plot the Bloch vector length $b = \sqrt{b_x^2+b_y^2+b_z^2}$ with $b_i$ the Bloch vector component in direction $i$. 

Over the measured range, the eigenstates show little decay. The decay of the superposition states is fitted with the function $f(x) = A e^{-(x/N_{1/e})^n}$. The fitted parameters can be found in Table~\ref{tab:mem_fit_param}. 

The use of the decoupling pulse $\pi_M$ on the memory qubit increases the $N_{1/e}$ by more than a factor $6$. Moreover, the initial Bloch vector length $A$ is higher with the $\pi_M$ pulse. This is mainly explained by the second round of phase stabilization~\cite{pompili_realization_2021} in between swapping the state onto the memory qubit and starting the entanglement generation process. The phase stabilization takes $\approx$350$\mu$ s and during this time the memory qubit is subject to intrinsic $T_{2}^{\ast}$ dephasing, which can be efficiently decoupled using the $\pi_M$ pulse.

\section{Communication qubit coherence}

In various parts of the protocol we decouple the communication qubits from the spin bath environment to extend their coherence time. On Alice, we start the decoupling when the first entangled link is established and stop when the results of the Bell-state measurement to teleport the state are sent by Charlie. On Bob, we decouple the communication qubit when the memory qubit is being re-phased. On Charlie, the communication qubit is decoupled from the point that entanglement with Bob is heralded up to the point where Bob has finished the Bell-state measurement, performed the CR check and has communicated the results. All these decoupling times are dependent on how many entanglement attempts are needed to generate the entangled link between Bob and Charlie. 

\begin{figure}
	\centering
	\includegraphics[width=0.8\linewidth]{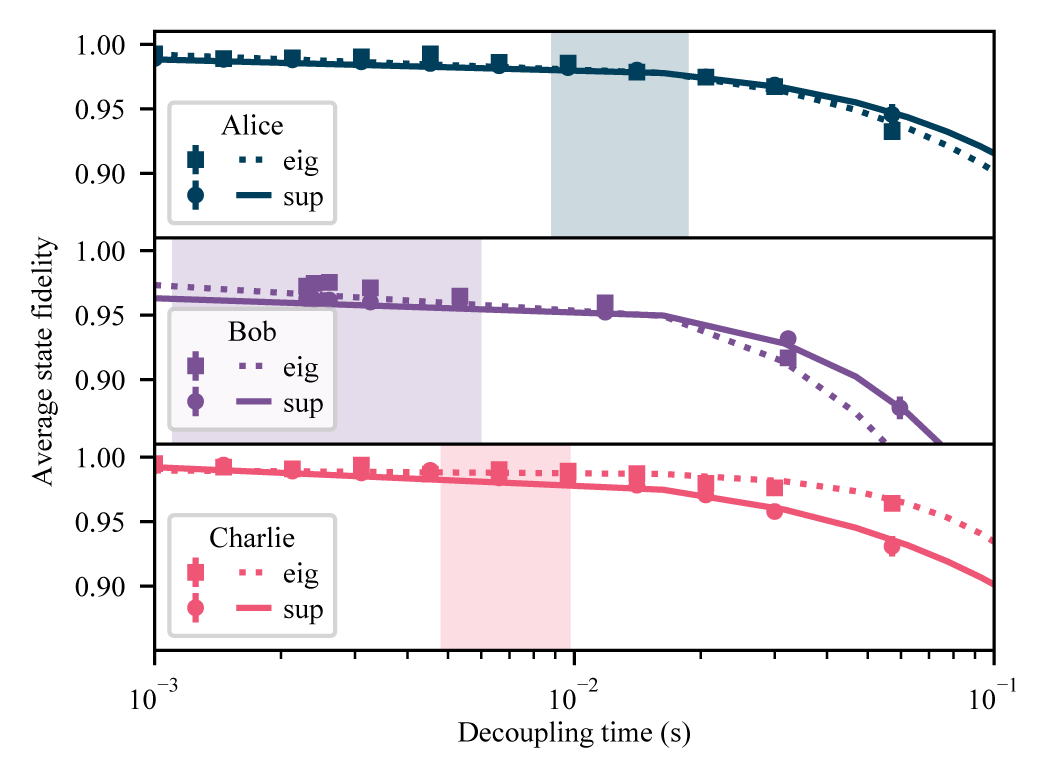}
	\caption{\label{fig:supp_el_decoupling} Decoupling of the communication qubits. The average state fidelity is plotted for different decoupling times for each setup. The shaded area represent the decoupling times used in the teleportation protocol.}
\end{figure}

We characterize the average state fidelities for different decoupling times, see Figure~\ref{fig:supp_el_decoupling}. We investigate eigenstates and superposition states separately. We fit the fidelity with the function $f(t) = A e^{-(t/\tau_{coh})^n} + 0.5$. The fitted parameters are summarized in Table~\ref{tab:el_dec_fit_params}. For each setup, the minimum and maximum used decoupling times are indicated by the shaded regions in Figure \ref{fig:supp_el_decoupling}. The left-most border is the decoupling time when the first entanglement attempt on Bob and Charlie would be successful, the right-most border when the last attempt before the timeout of $1000$ attempts would herald the entangled state.

\section{Basis-alternating repetitive readout}

\begin{figure}
	\centering
	\includegraphics[width=\linewidth]{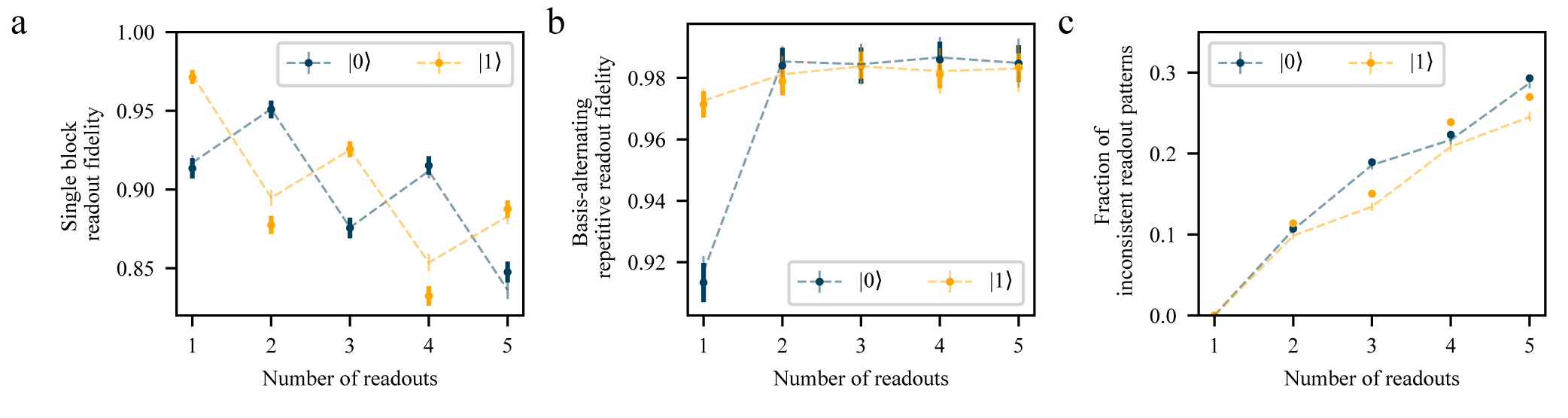}
	\caption{\label{fig:supp_rep_readout_charlie} Basis-alternating repetitive (BAR) readout results for Charlie's memory qubit. \textbf{a} Measured fraction of memory qubit states that were assigned 0 per readout block, for initialization in $\ket{0}$ and in $\ket{1}$.  \textbf{b} Readout fidelity of the basis-alternating repetitive readout scheme for different number of readout repetitions.  \textbf{c} Fraction of inconsistent readout patterns for different number of readout repetitions.  The dashed lines represent a numerical model using measured parameters.}
\end{figure}

In the main text we discuss the basis-alternating repetitive readout and the results on Bob's memory qubit are shown in Figure 3. Here we show the results for Charlie's memory qubit. We assign the state using the first readout and only accept the result when the consecutive readouts give a consistent pattern. The results for two different initial states of the memory qubit are plotted in Figure \ref{fig:supp_rep_readout_charlie}. We model the expected performance with a Monte Carlo simulation which takes into account the electron readout fidelities, the initial state populations and gate errors, see \url{https://doi.org/10.4121/16645969} \footnote{to be published}.
In the case of unconditional teleportation, the state is assigned using the first readout and is accepted regardless of the second readout result. 

\section{Teleportation results}
The numerical values of the data displayed in Figures 4b and 4c in the main text can be found in Tables~\ref{tab:results_tel} and ~\ref{tab:results_tel_bsm}, respectively. 

\section{Data acquisition and experimental rates}

\begin{figure}
	\centering
	\includegraphics[width=0.6\linewidth]{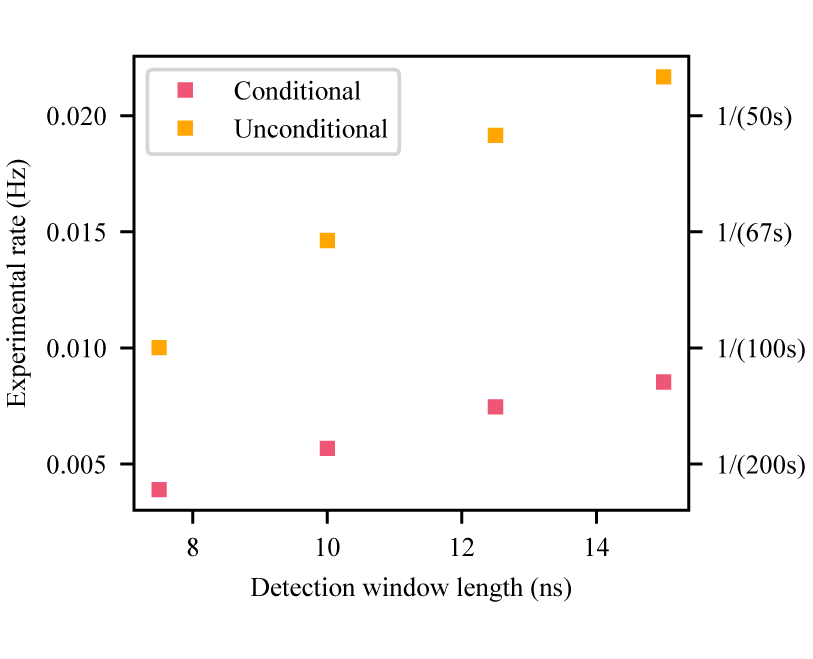}
	\caption{\label{fig:supp_exp_rates} Experimental rates of the conditional and unconditional teleportation protocol for different detection window lengths in the two-node entanglement generation. }
\end{figure}

For the data acquisition, we interleaved blocks of measurements with calibrations. We collected the data in blocks of $\approx$ 1 hour. In total, we have acquired 79 blocks of data, and we measured 2272 events ($\ket{+X} 382$ , $\ket{-X}  385$, $\ket{+Y} 385$  , $\ket{-Y} 378$ , $\ket{+Z} 375$  , $\ket{-Z} 367$) for the conditional teleportation over a time span of 21 days. We can determine the experimental rate including all overhead (such as CR checks, communication time and phase stabilization) by dividing the number of measured data points by the total measurement time. In Figure \ref{fig:supp_exp_rates} we plot the experimental rate for both the conditional and unconditional teleportation sequence. In the case of the unconditional teleportation, we accept all Bell-state measurement outcomes on Charlie and therefore the experimental rate is higher. For shorter detection windows during the two-node entanglement, the success probability per attempt is smaller and thus the experimental rate is lower. 

\section{Model of the teleported state}
A detailed model of the teleported state can be found at \url{https://doi.org/10.4121/16645969} \footnote{to be published}. The model comprises elements from~\cite{pompili_realization_2021} and is further extended for the teleportation protocol. We take the following noise sources into account 
\begin{itemize}
    \item imperfect Bell states between Alice and Bob, and between Bob and Charlie,
    \item dephasing of the memory qubit of Bob during entanglement generation between Bob and Charlie,
    \item depolarizing noise on the memory qubits of Bob and Charlie, due to imperfect initialization and swap gates,
    \item readout errors on the communication qubits of Bob and Charlie and readouts errors on the memory qubits of Bob and Charlie when using the basis-alternating readout scheme which result in incorrect feed-forward gate operations after the Bell-state measurements, 
    \item depolarizing noise on Alice during the decoupling sequence,
    \item ionization probability on Alice.
\end{itemize}

An overview of the input parameters and the effect of the different error sources is given in Tables~\ref{tab:error_budget_epr} and~\ref{tab:error_budget_teleportation}. 

\section{Effect of the 3 key innovations on the teleported state fidelity and experimental rate}
We assess the effect of each innovation on the teleportation protocol. First, we estimate the average state fidelity and experimental rate with a set of baseline parameters based on the performance in~\cite{pompili_realization_2021}. We use a timeout of $1000$ entanglement attempts for the second link (between Bob and Charlie) before aborting the protocol and starting over. In both Bell-state measurements, we continue the sequence for the outcomes "00" and "01" (communication qubit, memory qubit), or abort and start over (in the case of conditional teleportation). Then we incrementally add (1) the basis-alternating repetitive readout scheme for the memory qubits,(2) the improved memory qubit coherence and (3) the tailored heralding scheme of the remote entanglement generation. The results are summarized in \ref{tab:improvements}. 

\section{Estimated fidelity of state to be teleported}
The state to be teleported is prepared on the communication qubit of Charlie. Errors in the preparation originate from imperfect initialization and imperfect MW pulses, which are estimated to be $p_{init} = 1.2 \times 10^{-3}$ and $p_{MW} = 8 \times 10^{-3}$~\cite{kalb_dephasing_2018}. Averaged over the six cardinal states, we estimate the state preparation fidelity to be $\approx$ 0.995. 

\section{Calculation of teleported state fidelity without feed-forward operation}
In figure 4c in the main text we show the fidelity of the teleported state in case no feed-forward operations would have been applied on Alice. To extract this data we follow the same method as in~\cite{pfaff_unconditional_2014}. We perform classical bit flips on the measurement outcomes to counteract the effect of the feed-forward gate operations (as if the gate was not applied) for each Bell-state measurement outcome.  We do this for all six cardinal states and compute the average fidelity. We assume the errors of the gate in the feed-forward operations to be small.

\clearpage

\clearpage
\begin{table}[]
 \centering
     \caption{Memory qubit characteristics. In each setup we use a magnetic field with strength $B_z$ aligned to the NV axis. The nuclear spin precession frequencies ($\omega_{m_s = 0}$ and $\omega_{m_s = -1}$) depends on the electron spin state. From the frequency difference, the parallel component $A_\parallel$ of the hyperfine interaction can be estimated. Conditional (unconditional) pulses are applied by doing $N_{con}$ ($N_{unc}$) pulses on the electron spin with an inter-pulse delay of $\tau_{con}$ ($\tau_{unc}$).}
    \begin{tabular}{|c|c|c|c|c|}
    \hline
     Setup    & $B_z$  & $\omega_{m_s = 0}$ & $\omega_{m_s = -1}$ & $A_{\parallel}$ \\ \hline
     Bob  & 1890 Gauss  & $2\pi \times 2025$ kHz & $2\pi \times 2056$ kHz & $2\pi \times 30$ kHz \\ \hline
     Charlie & 165 Gauss  & $2\pi \times 177$ kHz & $2\pi \times 240$ kHz & $2\pi \times 63$ kHz \\ \hline
     \end{tabular}
     \newline
     \vspace{0.5cm}
     \newline
    \begin{tabular}{|c|c|c|c|c|}
    \hline
    Setup & $\tau_{con}$ & $N_{con}$ & $\tau_{unc}$ & $N_{unc}$\\ \hline
    Bob & 2.818 $\mu$s & 54 & 4.165 $\mu$s & 144 \\ \hline
    Charlie &6.003 $\mu$s & 56 & 11.996 $\mu$s & 30  \\ \hline
    \end{tabular}

    \label{tab:c13_couplings}
\end{table}

\begin{table}[]
\centering
\caption{ Estimated probabilities for the double optical excitation error and the double $\ket{0}$ occupancy error per node (values in percent).}
\begin{tabular}{|lll|}
\hline
Node              & Double optical excitation probability         & Double $\ket{0} $ occupancy probability           \\
\hline
Alice              & 4.1 $\pm$ 0.5 & 7.6  $\pm$ 0.4 \\
Bob (with Alice)   & 2.6  $\pm$ 0.3 & 4.9  $\pm$ 0.3 \\
Bob (with Charlie) & 6.9  $\pm$ 0.8 & 4.7  $\pm$ 0.8 \\
Charlie            & 5.7  $\pm$ 0.4 & 9.4  $\pm$ 0.4 \\ \hline
\end{tabular}
\label{tab:alpha_pde}
\end{table}

\begin{table}
\centering
\begin{tabular}{|p{2cm}|p{12 cm}|}
\hline
Parameter & Description  \\ \hline
$\gamma$ & Spontaneous emission rate of the excited state. \\ \hline
$\Omega(t)$ & Optical driving strength. \\ \hline
$\alpha$ & Initial population of the $\ket 0$ state. \\ \hline
$P_0$ & Probability of emitting 0 photons (ZPL or PSB). \\ \hline
$P_1$ & Probability of emitting 1 photons (ZPL or PSB).\\ \hline
$P_2$ & Probability of emitting 2 photons (ZPL or PSB or both).\\ \hline
$P_z$ & Probability that an emitted photon is a ZPL photon.\\ \hline
$P_{dz,1}$ & Probability that a ZPL photon is within the ZPL detection window, conditioned on a single ZPL photon being emitted. \\ \hline
$P_{db,1}$ & Probability that a PSB photon is within the PSB detection window, conditioned on a single PSB photon being emitted. \\ \hline
$P_{dz,2}$ & Probability that 2 ZPL photons are within the ZPL detection window, conditioned on two ZPL photons being emitted. \\ \hline
$P_{dz,3}$ & Probability that one ZPL photons is within the ZPL detection window and one is not, conditioned on two ZPL photons being emitted. \\ \hline
$P_{db,2}$ & Probability that 2 PSB photons are within the PSB detection window, conditioned on two PSB photon being emitted. \\ \hline
$P_{dz,3}$ & Probability that one PSB photons is within the PSB detection window and one is not, conditioned on two PSB photons being emitted. \\ \hline
$P_{dzb,1}$ & Probability that a ZPL photon is within the ZPL detection window and a PSB photon is within the PSB detection window, conditioned on one ZPL and one PSB photon being emitted. \\ \hline
$P_{dzb,2}$ & Probability that a ZPL photon is not within the ZPL detection window and a PSB photon is within the PSB detection window, conditioned on one ZPL and one PSB photon being emitted. \\ \hline
$P_{dzb,3}$ & Probability that a ZPL photon is within the ZPL detection window and a PSB photon is not within the PSB detection window, conditioned on one ZPL and one PSB photon being emitte. \\ \hline
$\eta_z$ & Total transmission and detection efficiency of ZPL photons. \\ \hline
$\eta_p$  & Total transmission and detection efficiency of PSB photons. 
 \\ \hline
\end{tabular}
\caption{Explanation of the parameters used in the numerical simulation of the entanglement generation protocol. }
\label{tab:tableS1}
\end{table}

\clearpage
\begin{table}[h]
    \centering
    \begin{tabular}{|llll|}
    \hline
    &      A &     $N_{1/e}$  &     n    \\ \hline
    With ent. att. with $\pi_M$   & 0.875 $\pm$  0.015 &  5327 $\pm$  319 &  1.13 $\pm$  0.11   \\
    With ent. att. without $\pi_M$  & 0.806 $\pm$  0.019 &   848 $\pm$   39 &  1.21 $\pm$  0.09   \\
    Without ent. att. with $\pi_m$  & 0.884 $\pm$  0.011 &  5239 $\pm$  163 &  1.94 $\pm$  0.16   \\
    Without ent. att. without $\pi_M$ & 0.807 $\pm$  0.019 &   880 $\pm$   34 &  1.37 $\pm$  0.10   \\
    \hline
    \end{tabular}
    \caption{Fitted parameters for the memory coherence decay of the superposition states.}
    \label{tab:mem_fit_param}
\end{table}
\pagebreak

\begin{table}[h]
    \centering
        \begin{tabular}{|lllll|}
        \hline
        {}& {}&      A   &    $\tau_{coh} (s)$ &     n \\
        \hline
        Alice & Eigenstate      & 0.4930 $\pm$ 0.0013 &  0.459 $\pm$ 0.012 &  1.04 $\pm$  0.03  \\
        {}   &  Superposition   & 0.4889 $\pm$ 0.0018 &  0.54 $\pm$ 0.02  &  1.07 $\pm$  0.05   \\
        Bob &   Eigenstate      & 0.4738 $\pm$ 0.0011 &  0.130 $\pm$ 0.003 &  1.41 $\pm$  0.04  \\
        {}  &   Superposition   & 0.4634 $\pm$ 0.0015 &  0.177 $\pm$ 0.006 &  1.47 $\pm$  0.06   \\
        Charlie&Eigenstate      & 0.4897 $\pm$ 0.0009 &  0.357 $\pm$ 0.007 &  1.67 $\pm$  0.06  \\
        {} &    Superposition   & 0.4936 $\pm$ 0.0019 &  0.56 $\pm$ 0.02  &  0.92 $\pm$  0.04   \\
        \hline
        \end{tabular}
    \caption{Fitted parameters for average state fidelity state during communication qubit decoupling.}
    \label{tab:el_dec_fit_params}
\end{table}
\pagebreak

\clearpage

\begin{table}[]
    \centering
    \caption{Numerical values of the data displayed in Figure 4b of the main text. }
\begin{tabular}{|lr|}
\hline
{} &   Teleported state fidelity \\
\hline
X   & 0.760 $\pm$  0.024 \\
-X  & 0.745 $\pm$  0.025 \\
Y   & 0.656 $\pm$  0.027 \\
-Y  & 0.651 $\pm$  0.027 \\
Z   & 0.731 $\pm$  0.026 \\
-Z  & 0.671 $\pm$  0.027 \\
Average & 0.702 $\pm$  0.011 \\
\hline
\end{tabular}
    \label{tab:results_tel}
\end{table}

\begin{table}[]
    \centering
        \caption{Numerical values of the data displayed in Figure 4c of the main text.}
\begin{tabular}{|cr|}
\hline
\begin{tabular}{l}Bell-state measurement outcome \\ (memory qubit, communication qubit)\end{tabular} &   Average teleported state fidelity  \\
\hline
00              & 0.707 $\pm$  0.015 \\
01              & 0.696 $\pm$  0.014 \\
10              & 0.698 $\pm$  0.015 \\
11              & 0.671 $\pm$  0.014 \\
No feed forward & 0.501 $\pm$  0.007 \\
\hline
\end{tabular}

    \label{tab:results_tel_bsm}
\end{table}

\clearpage

\clearpage

\begin{table}[h]
    \caption{Overview of parameters used in the simulations for the two-node entangled states. The error due to the $\ket 0$ state populations is a result of the single click protocol. For the other error sources we compute the estimated infidelity as if it was the only error source present apart from the protocol error. This allows easy comparison between the different error sources.}
    \centering
        \begin{tabular}{|lrrrr|}
        \hline
        {}  &  \small{Parameter AB} &  \small{Parameter BC} &    \small{Infidelity $\Psi_{AB}$}  &  \small{Infidelity $\Psi_{BC}$} \\
        \hline
        Detection window length         &  15 ns &  15 ns & {} & {} \\
        Detection probability setup 1   &  3.4$\times 10^{-4}$ &  4.3$\times 10^{-4}$ & {} & {} \\
        Detection probability setup 2   &  5.1$\times 10^{-4}$ &  2.4$\times 10^{-4}$ & {} & {} \\
        Average detection probability PSB &  0.10 &  0.12 & {} & {} \\
        $\ket 0$ state populations $(\alpha_1, \alpha_2)$        &(0.07, 0.05) &(0.05, 0.1) & 5.5 $\times 10^{-2}$ &  6.7 $\times 10^{-2}$ \\
        Dark count rate                 &     10 Hz &     10 Hz &  5.1 $\times 10^{-3}$ &  5.3 $\times 10^{-3}$ \\
        Visibility      &      0.90 &      0.90 &  2.4 $\times 10^{-2}$ &  2.4 $\times 10^{-2}$ \\
        Average double excitation probability& 0.06 &      0.08 &  5.5 $\times 10^{-2}$ &  7.1 $\times 10^{-2}$ \\
        Optical phase uncertainty               &     21$^o$&     12$^o$ & 3.1 $\times 10^{-2}$ &  1.0 $\times 10^{-2}$ \\ \hline
        All error sources combined  & {} & { }& 0.16 &  0.17 \\
        \hline
        \end{tabular}

    \label{tab:error_budget_epr}
\end{table}

\begin{table}[h]
    \caption{Overview of parameters used in the simulations for the average teleported state fidelity in case of a conditional Bell-state measurement on Charlie. For each error sources we compute the estimated infidelity as if it was the only error source present apart from the single click protocol errors of the two-node entangled states. This allows easy comparison between the different error sources.}
    \centering
        \begin{tabular}{|lrl|}
        \hline
        {}  &                    Parameter &       Infidelity \\
        \hline
        Ionization probability Alice                &       0.7$\%$                 &  0.6 $\times 10^{-2}$ \\
        Depolarizing noise Alice                    &       0.04                    &  1.7 $\times 10^{-2}$ \\
        Depolarizing noise memory qubit Bob         &       0.12                    &  5.0 $\times 10^{-2}$ \\
        Dephasing noise memory qubit Bob ($N_{1/e}, n$)&    (5300, 1.1)             &  2.1 $\times 10^{-2}$ \\
        Depolarizing noise memory qubit Charlie         &       0.14                &  5.9 $\times 10^{-2}$ \\
        Readout fidelities memory qubit Bob  $(\ket{0},\ket{1})$ & (0.99, 0.99)   &  0.6 $\times 10^{-2}$ \\
        Readout fidelities communication qubit Bob $(\ket{0},\ket{1})$&(0.93, 0.995)&  0.3 $\times 10^{-2}$ \\
        Readout fidelities memory qubit Charlie $(\ket{0},\ket{1})$ &(0.98, 0.98) &  1.1 $\times 10^{-2}$ \\
        Readout fidelities communication qubit Charlie $(\ket{0},\ket{1})$&(0.92, 0.99) &  0.6 $\times 10^{-2}$ \\ \hline
        Two-node entangled states combined &  {} &  0.192 \\
        All error sources combined                  &  {}  &  0.305 \\
        \hline
        \end{tabular}

    \label{tab:error_budget_teleportation}
\end{table}

\begin{table}[h]
\centering
    \caption{Simulated effect of the innovations on the teleported state fidelity and experimental rate.}
\begin{tabular}{|p{105mm}|l|l|}
\hline                                                  & Fidelity   & Rate (Hz)  \\\hline
Baseline parameters using timeout = 1000, BSM outcomes (communication qubit, memory qubit) = "00" or "01" & 0.666 & 1/(53s)  \\\hline
With basis-alternating repetitive readout & 0.679 & 1/(73s)  \\\hline
With improved memory coherence     & 0.687 & 1/(73s)  \\\hline
With tailored heralding scheme            & 0.695 & 1/(74s) \\ \hline
\end{tabular}

    \label{tab:improvements}
\end{table}

\end{document}